\documentclass[prl,twocolumn,showpacs,superscriptaddress,floatfix, nofootinbib]{revtex4-1}
\usepackage[caption=false]{subfig}
\usepackage{amsmath,graphicx}
\usepackage{epsfig}
\usepackage{amsmath,amssymb,mathrsfs,esint}
\usepackage{gensymb,textcomp}
\usepackage[bookmarks=true,
   colorlinks=true,
   linkcolor=blue,
   urlcolor=blue,
   citecolor=blue,
   bookmarks=true,
   hyperindex=true
]{hyperref}
\usepackage{natbib}
\usepackage{relsize}
\usepackage{float}
\usepackage{dsfont}
\usepackage{lipsum}
\usepackage{multibib}
\newcites{supp}{Supplementary References}

\usepackage{bm}

\newcommand{\be}{\begin{equation}}
\newcommand{\ee}{\end{equation}}

\newcommand{\beq}{\begin{eqnarray}}
\newcommand{\eeq}{\end{eqnarray}}
\DeclareMathOperator{\Tr}{Tr}

\def\H1{\widehat{H}_1}

\renewcommand{\i}{\ensuremath{\mathrm{i}}}

\newcommand{\ket}[1]{\left| #1 \right>}
\newcommand{\bra}[1]{\left< #1 \right|}

\renewcommand{\Re}[1]{ \text{Re} \left\{ #1\right\} }

\usepackage{soul}
\usepackage{cancel}

\usepackage[normalem]{ulem}
\begin{document}

\title{Driven-dissipative time crystalline phases \\ in a two-mode bosonic system with Kerr nonlinearity}

\author{L.R. Bakker}
\affiliation{Institute for Theoretical Physics, Universiteit van Amsterdam, Science Park 904, Amsterdam, The Netherlands}
\affiliation{Russian Quantum Center, Skolkovo, Moscow 143025, Russia}
\author{M.S. Bahovadinov}
\affiliation{Russian Quantum Center, Skolkovo, Moscow 143025, Russia}
\affiliation{Physics Department, National Research University Higher School of Economics, Moscow, 101000, Russia}

\author{D.V. Kurlov}
\affiliation{Russian Quantum Center, Skolkovo, Moscow 143025, Russia}

\author{V. Gritsev}
\affiliation{Institute for Theoretical Physics, Universiteit van Amsterdam, Science Park 904, Amsterdam, The Netherlands}
\affiliation{Russian Quantum Center, Skolkovo, Moscow 143025, Russia}

\author{Aleksey K. Fedorov}
\affiliation{Russian Quantum Center, Skolkovo, Moscow 143025, Russia}
\affiliation{National University of Science and Technology ``MISIS”,  Moscow 119049, Russia}

\author{Dmitry O. Krimer}
\affiliation{Institute for Theoretical Physics, Vienna University of Technology (TU Wien),
Wiedner Hauptstraße 8-10/136, A–1040 Vienna, Austria}

\begin{abstract}
For the driven-dissipative system of two coupled bosonic modes in a nonlinear cavity resonator, we demonstrate a sequence of phase transitions from a trivial steady state to two distinct dissipative time crystalline phases.
These effects are already anticipated at the level of the semiclassical analysis of the Lindblad equation using the theory of bifurcations and are further supported by the full quantum (numerical) treatment.
The system is predicted to exhibit different dynamical phases characterized by an oscillating non-equilibrium steady state with non-trivial periodicity, which is a hallmark of time crystals. We expect that these phases can be directly probed in various cavity QED experiments.
\end{abstract}

\maketitle
\textit{Introduction.} Nonlinear quantum optical effects are of great importance both for fundamental research and various applications, in particular in quantum information technologies~\cite{Imoto1985,Harris1990,Turchette1995,Kumar2010,Chang2014,doi:10.1063/5.0065222}. Realistic settings of quantum experiments require considering not only sizable nonlinear effects, but also an interplay between external driving and dissipation caused by the fundamentally open nature of such systems. A system of paramount importance is a driven-dissipative model of bosonic modes with the Kerr nonlinearity \cite{Zhang_2018,Xue_2019,kerr-rev,Tikan_2021,Englebert_2021}. For example, a qubit encoded in quantum harmonic oscillators~\cite{GKP2001} can be made stable against environment-induced decay using an interplay between Kerr-type interactions and squeezing~\cite{Grimm2020,Yurke1986,Kirchmair2013}. On the fundamental side, non-equilibrium bosonic systems with a Kerr nonlinearity may exhibit novel dynamical phases, such as time crystals~\cite{Alaeian2021-1,Alaeian2021-2,Cristobal2019}. 

The time crystal (TC) phase of matter has been predicted theoretically in isolated Floquet driven systems and driven-dissipative systems~\cite{Ciuti2017,Seibold2020,Muniz2020,Cristobal2020,Roberts2020,Gong2018, Ippoliti2021,Cosme2019,Ke_ler_2020,Alaeian2022,Buca2019-2,Buca2019-1, Kelly2021} and has recently been observed experimentally \cite{Yao2018,Zhang2017,Choi2017,Mi2022,Keler2021,Dogra2019}.
Time crystals were originally introduced as the temporal analogue of spatial crystals where the time (rather than spatial) translation symmetry of a system is broken~\cite{Wilczek2012}. Crucially, the time crystalline phase of matter would then be resistant to entropy increase ~\cite{Yao2018,khemani2019, Sacha_2017,Else2020}. This property makes the TC phase of matter an interesting candidate for quantum hardware devices, where entropy growth and spontaneous decay leads to corruption of stored information. 

In this work we demonstrate that a system of two driven-dissipative coupled bosonic modes that are trapped in an optical cavity with Markovian dissipation exhibits intriguing dynamical behavior featuring inter alia time-crystalline phases. In the semiclassical regime, the system is shown to undergo a series of sub- and supercritical Hopf bifurcations between different stationary solutions. The Hopf bifurcations are responsible for the periodic dynamics emerging in the form of limit cycles in the phase space of a system~\cite{Glend} - a phenomenon that is absent in a single-mode bosonic system
with a Kerr nonlinearity \cite{Drummond1980,Bartolo2016}. 
\begin{figure}[t]
    \centering
    \includegraphics[width = 8cm]{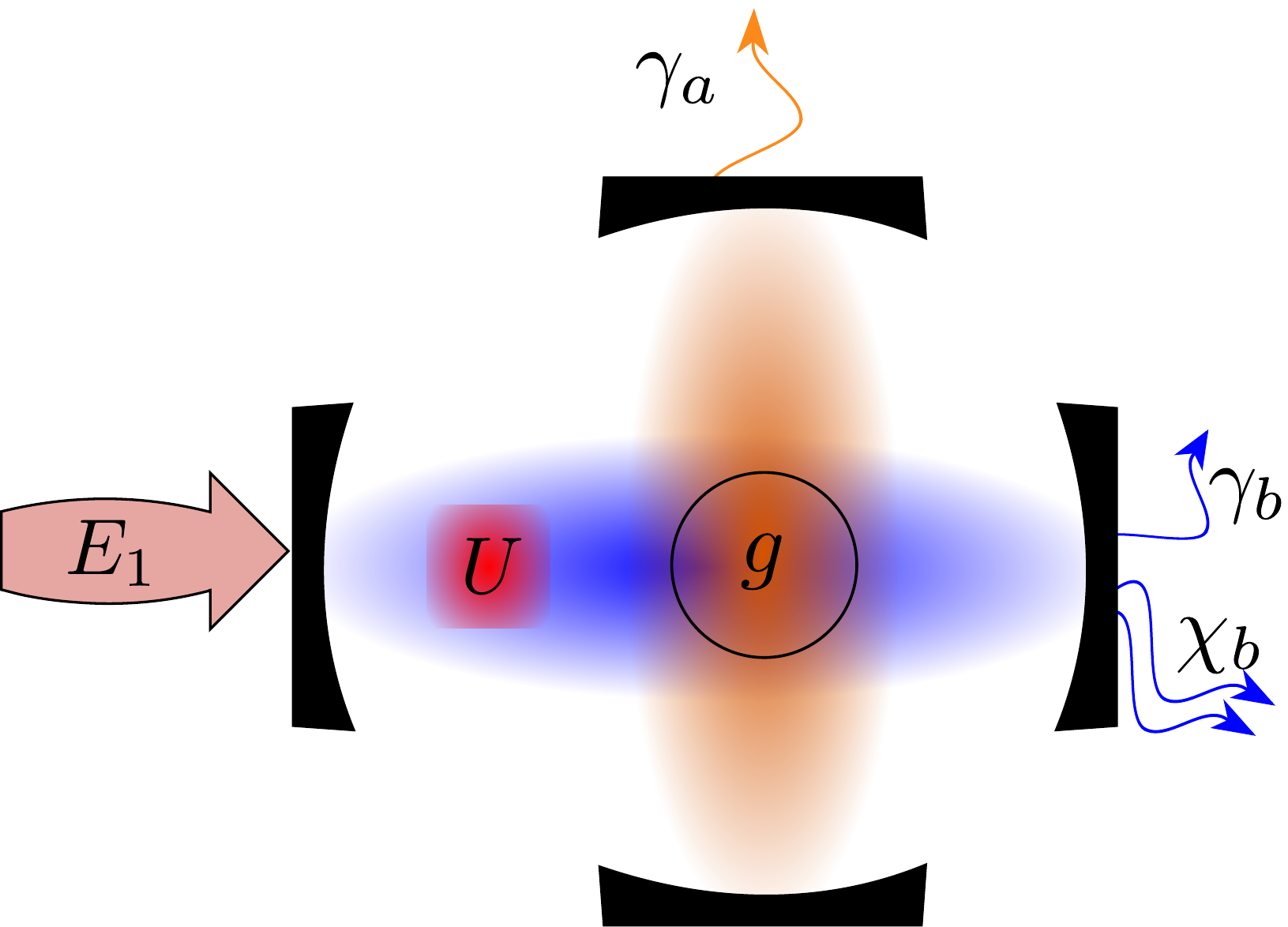}
\caption{Schematic representation of our setup: A cavity with two bosonic modes $a$ (orange cloud) and $b$ (blue cloud) coupled with strength $g$ to each other, see Eq. \eqref{Lindblad_eq_ab}. A Kerr interaction with strength $U$ is generated by a nonlinear element (red square) for the $b$-mode. The cavity is driven coherently by a single photon drive $E_1$. The cavities allow for the decay of the modes with the single-photon rates $\gamma_{a,b}$ and two-photon rate $\chi_b$ (orange and blue arrows).}
    \label{fig:System_Pictorial}
\end{figure}
Most importantly, we find a period doubling behavior suggesting existence of multiple distinct, non-trivial TC phases present in the system.
The presence of the limit cycles on the semiclassical level can be considered as an indicator for possible (dissipative) TC phases in the full quantum dynamics of our system. Indeed, in the quantum regime we observe signatures of multiple nonequilibrium phase transitions in the form of the closure of the dissipative gap in the Liouvillian spectrum. We provide evidence that the semiclassical predictions are in many aspects consistent with the results obtained in the framework of the full quantum mechanical approach.

Our analytical approach is based on a combination of the Lie-algebraic disentanglement technique ~\cite{Gritsev2017,Ringel_2013,Bakker2020,Charzyski2013,Wei1963,Wei1964,scully_zubairy_1997} and a semiclassical approximation (see Supplementary material for detailed exposition).
The results in the quantum regime are found using exact diagonalization (ED) methods and by performing Monte Carlo simulations for trajectories of observables. The Monte Carlo simulations are not as sensitive to system size scaling as ED computations and therefore allow us to investigate larger system sizes. Using a combination of all aforementioned methods we conclude that different time crystalline phases exist in a broad range of values of the single-photon driving amplitude.

\textit{The model.} We consider two driven-dissipative coupled modes in a cavity~\cite{Walls2006} (see Fig.\,\ref{fig:System_Pictorial}) described by the following Hamiltonian ($\hbar=1$):
\be \label{H_micro}
\begin{aligned}
	\hat H &= \omega_a \hat a^{\dag}\hat a + g \hat b^{\dag} \hat a + g^* \hat b \hat a^{\dag} + \omega_b \hat b^{\dag} \hat b\\
		  & + E_1(t) \hat b + E_1^*(t)\, \hat b^{\dag} + \frac{U}{2} \hat b^{\dag}\hat b^{\dag}\hat b \hat b,
\end{aligned}
\ee
where $\hat a$, $\hat b$ ($\hat a^{\dag}$, $\hat b^{\dag}$) are bosonic annihilation (creation) operators. Parameters $\omega_{j} > 0$ are the cavity frequencies of the $a$- and $b$-modes, $g$ is the coupling strength between the modes, $E_1(t)$ determines the driving protocol of the $b$ mode and $U$ is the Kerr interaction strength. In nonlinear media $U\sim n_{2}\omega_{0}^{2}/(n_{0}^{2}V_{eff})$, where $n_{0,2}$ are linear and nonlinear refractive indexes, $\omega_{0},V_{eff}$ are the mode frequency and effective volume respectively. Both $a$- and $b$-modes are coupled to a zero-temperature Markovian environment. The $a$-mode experiences only the single-photon losses, whereas the $b$-mode is prone to both single- and two-photon losses \cite{note1}.
The overall time evolution of the system is then governed by the Lindblad equation \cite{Lindblad1976},
\be \label{Lindblad_eq_ab}
	\dot \rho = - i[\hat H,\rho] + \frac{\gamma_a}{2}{\cal D}[\hat a]\rho + \frac{\gamma_b}{2}{\cal D}[\hat b]\rho + \frac{\chi_b}{2}{\cal D}[\hat b^2]\rho \equiv {\cal L}\rho,
\ee
where
${\cal D}{[\hat L]} \rho = 2 \hat L \rho \hat L^{\dag} - \hat L^{\dag} \hat L \rho - \rho \hat L^{\dag} \hat L$ is the dissipator, $\hat H$ is given by Eq. (\ref{H_micro}), and ${\cal L}$ is the Liouvillian. Moreover, $\gamma_{j} > 0$ and $\chi_j > 0$ represent the cavity single and double mode loss rates, correspondingly.

\textit{Semiclassical analysis.} 
In the semiclassical approximation, the Lindblad equation is reduced to the master equation (see supplemental information for details)
\be \label{CS_res_eqs_autonomous_vec}
	\dot {\boldsymbol \xi}(t) = {\boldsymbol A}\left( |z|^2 \right) {\boldsymbol \xi}(t) + {\boldsymbol \eta},
\ee
where the matrix ${\boldsymbol A}$ reads as
\be 
	{\boldsymbol A}\left( |z|^2 \right) = \left(
	\begin{array}{cccc}
		\tilde \kappa_a & -i g^* &0 & 0 \\
		-i g & \varphi\!\left( |z|^2\right) & 0 & 0 \\
		0 & 0 & \tilde \kappa_a^* & i g \\
		0 & 0 & i g^* & \varphi^*\!\left( |z|^2\right) \\
	\end{array}
	\right).
\ee
Here the vectors ${\boldsymbol \xi}( y, z, y^*, z^* )^T$ and $\boldsymbol{\eta}( 0, - i {\cal E}_1^*, 0,  i {\cal E}_1 )^T$ are defined by $y(t) = \exp(\tilde{\kappa}_a t)\Tr[\hat{a}\rho(t)]$ and $z(t) = e^{i\omega_1 t}b(t)$, for a periodic drive $E_1(t) = {\cal E}_1e^{i\omega_1 t}$ and $\varphi\!\left( |z|^2 \right) = \tilde \kappa_b + K |z|^2$, $\tilde \kappa_j = \kappa_j + i \omega_1 = -i \Delta_j - \gamma_j/2$, $\Delta_j = \omega_j - \omega_1$ for $j= a,b$, and $K = -\chi_b - i U$.

\begin{figure*}[ht!]
    \centering
    \includegraphics[width = \textwidth]{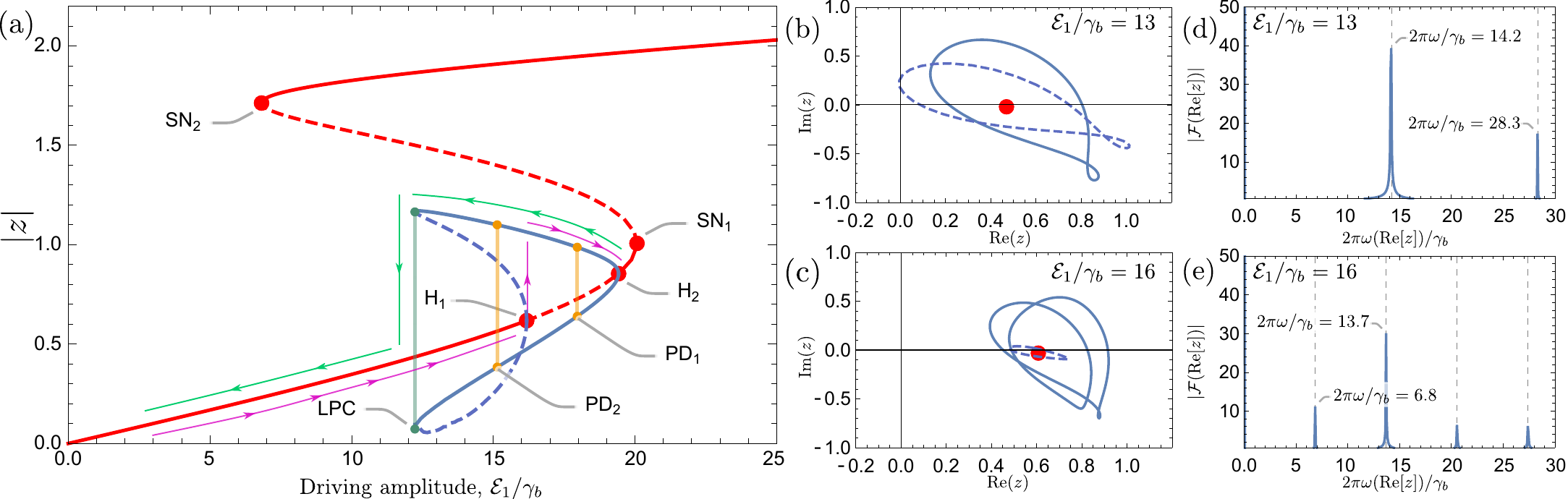}
    \caption{(a) Solid (dashed) red curve depicts the semiclassical stable (unstable) steady state solution for the b-mode particle number, $\sqrt{n_b} = |z|$, as a function of the driving amplitude $\mathcal{E}_1$. The points $\text{H}_1$ and $\text{H}_2$ are sub- and supercritical Hopf bifurcations, respectively. Stable (Unstable) limit cycles emerging from $H_2$ ($H_1$) are depicted using the solid (dashed) blue lines, indicating the absolute values in between which the oscillations occur. Stable and unstable limit cycles annihilate at the Limit Point cycle (LPC). $SN_{1,2}$ are saddle-node points of the optical bistability. $\text{PD}_1$ ($\text{PD}_2$) corresponds to a period doubling bifurcation when passing this point from upper (lower) values of $\mathcal{E}_1$. Between the points $\text{PD}_1$ and $\text{PD}_2$ is the region of limit cycles with a double loop structure as shown in (c). The magenta and green arrows indicate the forward and backward sweeping trajectories as outlined in the main text. (b), (c): examples of normal and period doubled limit cycles (solid blue line) and unstable limit cycle (dashed blue line) in the ($\text{Re}[z]$, $\text{Im}[z]$)-plane. The red dot represents a stationary state lying on the lower red curve from (a). (d) and (e) show the Fourier spectrum of $\text{Re}[z]$ when the dynamics is represented by stable limit cycles shown in (b) and (c). The system parameters are $\gamma_a = \chi_b = 1$, $g = U = \Delta_a = 10$ and $\Delta_b = -20$ (computed in units of $\gamma_b$). Data for limit cycles was partially obtained using the MatCont package \cite{matcont}.}
    \label{fig:sc_analysis}
\end{figure*}

\textit{Results.} We proceed with the analysis of the semiclassical steady-state solution by setting $\dot{\boldsymbol{\xi}} = 0$ into Eq.\eqref{CS_res_eqs_autonomous_vec} and obtain an S-shape [see solid and dashed red curves in Fig. \ref{fig:sc_analysis}(a)] that is well-known in the context of systems with Kerr-type interactions \cite{Drummond1980, Bartolo2016, Krimer2019}.
\begin{figure*}[t]
    \centering
    \includegraphics[width = \textwidth]{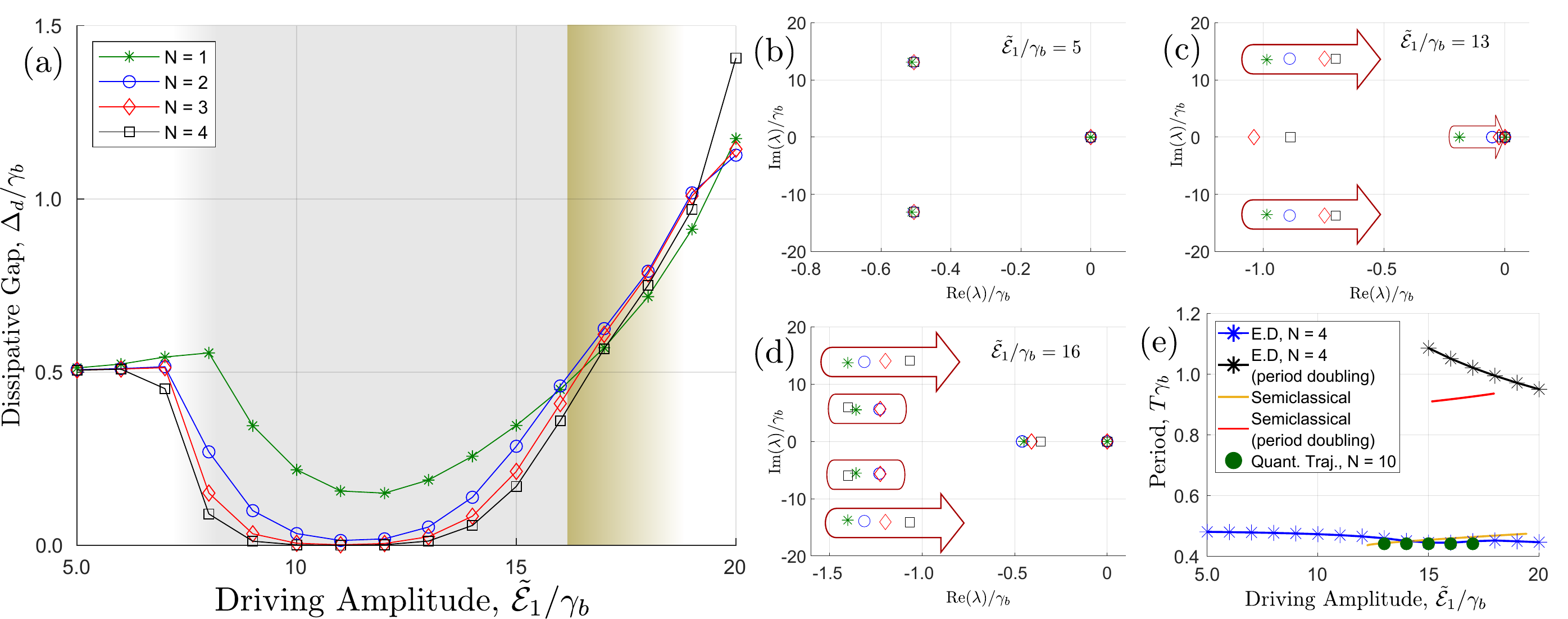}
    \caption{(a) The dissipative gap as a function of the driving amplitude $\tilde{\mathcal{E}}_1/\gamma_b$.
    Qualitatively different phases are colored in by hand, based on the underlying gap closure behaviors. (b) Eigenvalues of the first few decaying modes $\lambda_i$ for $\tilde{\mathcal{E}}_1/\gamma_b = 5$ and different values of $N$. Here, the gap does not close as we increase $N$. (c) The spectrum for $\tilde{\mathcal{E}}_1/\gamma_b = 13$. The arrows enclosing symbols depict how the eigenvalues converge towards the imaginary axis ($Re(\lambda_i)=0$) as a function of $N$. (d) Spectrum for $\tilde{\mathcal{E}}_1/\gamma_b = 16$. An additional eigenvalue enclosed in rectangles features gap closing behavior with increasing $N$, having an imaginary part that is about half as large as the original eigenvalues enclosed within the arrows. This eigenvalue resembles the period doubling found in the semiclassical case.
    (e) Periods of quantum oscillations, $T=2\pi/\text{Im}(\lambda_i)$, as a function of the driving amplitude. All system parameters are chosen the same as in Fig. \ref{fig:sc_analysis}. The results from the quantum trajectories are obtained by averaging 3000 trajectories for a duration of $t\gamma_b = 15$ with $\gamma_b dt = 0.005$ and using the software provided in~\cite{Qutip2}.}
    \label{fig:ED_analysis}
\end{figure*}
Looking more closely at the dynamical equations, however, we uncover several non-trivial system behaviors. The analysis of the dissipative system \eqref{CS_res_eqs_autonomous_vec} can be done using available tools developed for the theory of bifurcations \cite{Guckenheimer1983,matcont,Glend}.
In our system we find that the steady state solution has several interesting features, summarized in Fig.~\ref{fig:sc_analysis}(a). 
The steady state outside of the region of bistability (S-shape) is represented by a stable stationary solution. If we choose an initial state in this interval of $\mathcal{E}_1$ and let the system time evolve, it will relax to a respective stationary value on one of the (solid) red curves in Fig. \ref{fig:sc_analysis}(a). Most importantly, we find an interval of $\mathcal{E}_1$, where limit cycles are other possible time-dependent steady state solutions. In order to probe the systems behavior in more detail, we use the following approach. We start by considering a coherent drive $\mathcal{E}_1 = 0$, for which the steady state solution corresponds to zero particle number excitations in both the $b$- and $a$-modes, as expected. As we gradually increase the value of the coherent drive (for simplicity we assume $\arg(\mathcal{E}_1) = 0$), after some transient behavior the system will settle into a respective stationary solution for the $b$-expectation value, which lies on the lower solid red curve starting from $z=0$ in Fig.~\ref{fig:sc_analysis}(a). In this manner we can iteratively increase the driving amplitude, following the path outlined by the magenta arrow (arrow going left to right) in Fig.~\ref{fig:sc_analysis}(a). 

As we increase the driving amplitude to within the region of bistability, we encounter a region of instability between the points $\text{H}_1$ and $\text{H}_2$ (designated by a dashed curve in Fig.~\ref{fig:sc_analysis}(a) between the aforementioned points). The transition from stable to unstable solutions are accompanied by a subcritical Hopf bifurcation at $\text{H}_1$ and a supercritical Hopf bifurcation at $\text{H}_2$. Increasing the driving amplitude to beyond $\text{H}_1$, the system will jump to a limit cycle solution, so that the variable $z$ oscillates in time between some maximal and minimal values as designated by solid blue curves in this figure. The limit cycles for the driving amplitudes lying between the points $\text{PD}_2$ and $\text{PD}_1$ have qualitatively the same double-loop structure as exemplified in Fig.~\ref{fig:sc_analysis}(c) for $\mathcal{E}_1 = 1.6$. If one keeps increasing the driving amplitude, probing the limit cycle solutions for each value of ${\cal E}_1$, the limit cycles will half their period at the point $\text{PD}_1$ in Fig.~\ref{fig:sc_analysis}(a). As $\mathcal{E}_1$ increases further, its dimensions in the phase space decrease and eventually shrink to zero at the threshold point $\text{H}_2$, where the periodic solution ceases to exist (via the supercritical Hopf bifurcation). Above this threshold value, there is a stable stationary state between the points $\text{H}_2$ and $\text{SN}_1$. 

In addition, we disclose the bifurcation scenario following an inverse route by starting from the stationary state slightly below the saddle-node bifurcation, $\text{SN}_1$, and gradually decreasing the driving amplitude $\mathcal{E}_1$ [green path designated by arrows in Fig.~\ref{fig:sc_analysis}(a)]. This will lead to a partially different dynamical scenario associated with the hysteretic behavior shown in this figure. Specifically, when one decreases $\mathcal{E}_1$ below $\text{H}_2$, limit cycles are time-dependent steady states in the interval between $\text{H}_2$ and $\text{PD}_1$ as expected. Subsequently, the limit cycles double their period at $\text{PD}_1$. For driving amplitudes below $\text{H}_1$, unstable limit cycles are also possible solutions [dashed limit cycle in Fig.~\ref{fig:sc_analysis}(b)] that, however, can not be experimentally observed. At the point $\text{PD}_2$, the limit cycles half their period. With further decrease of $\mathcal{E}_1$, the stable and unstable limit cycles ultimately annihilate at the Limit Point Cycle (LPC). When decreasing the driving amplitude below the LPC point, the time-dependent steady state will jump down to a stationary state lying on the lower red curve.
Thus, on the semiclassical level, the system demonstrates a series of continuous and discontinuous phase transitions with hysteretic behavior.

In the next step, we compare the semiclassical results to the full quantum mechanical dynamics of the system. Using a representation of bosonic creation and annihilation operators in a truncated Fock basis, we can compute the spectrum of our system. The thermodynamic limit is reached when the driving amplitude approaches infinity, $\mathcal{E}_1 \rightarrow \infty$, while the product $F\sqrt{U}$ is kept fixed (the so-called `weak interaction limit')~\cite{Ciuti2017,Carmichael2015}. In addition, the product of $F\sqrt{\chi}$ should remain constant. We introduce a dimensionless parameter $N$ to keep track of the particle number and quantify the large $N$-limit as follows:
\be
\mathcal{E}_1 = \tilde{\mathcal{E}}_1\sqrt{N},\qquad U = \frac{\tilde{U}}{N}, \qquad \chi = \frac{\tilde{\chi}}{N}.
\ee
We obtain a qualitative picture of the quantum mechanical solution as a function of $N$ which is summarized in Fig.~\ref{fig:ED_analysis}. In general, the quantum mechanical results agree to a large extent with the semiclassical predictions. In the region of optical bistability shown in Fig.~\ref{fig:sc_analysis} ($7\lesssim \tilde{{\cal E}}_1/\gamma_b	\lesssim 20$), the dissipative gap, defined as the largest real part of the non-zero eigenvalues, closes rapidly, indicating a presence of a dissipative phase transition within this region. As we increase the driving amplitude to $\tilde{{\cal E}}_1/\gamma_b \approx 13$ (starting from $\tilde{{\cal E}}_1/\gamma_b \approx 7$), a pair of eigenvalues starts to approach the imaginary axis as a function of $N$ (see Fig.~\ref{fig:ED_analysis}(c)). 
The resulting quantum oscillations are the quantum mechanical analogue of the limit cycles observed in the semiclassical case and indicate the time-crystalline phase. This analogy between semiclassical and quantum oscillations can be derived by comparing the inverse of the imaginary parts of the eigenvalues responsible for quantum oscillations with the periods of limit cycles [see Fig.~\ref{fig:sc_analysis}(d), (e)]. When increasing the driving amplitude further to a value of $\tilde{{\cal E}}_1/\gamma_b \approx 16$ [Fig.~\ref{fig:ED_analysis}(d)], another set of eigenvalues (enclosed in rectangles) starts to approach the imaginary axis with approximately half the imaginary value of the modes that were observed before (enclosed in arrows). Thus, there is a strong indication of quantum behavior resembling the period doubling found in the semiclassical case that shows up in a similar parameter range. The appearance of the modes enclosed in rectangles then indicates a period doubled time crystalline phase. Unlike in the semiclassical case, however, we cannot find a signature of period \textit{halving} in the quantum regime as we increase the driving amplitude beyond $\tilde{\mathcal{E}}_1/\gamma_b = 16$. Rather, the upper bound on the driving amplitude for the system being in the period doubled time crystal phase remains undetermined. Furthermore, in the parameter regime $\tilde{\mathcal{E}}_1/\gamma_b \gtrsim 16$ the gap closure behavior is dominated by the oscillating (hard) modes rather than the (soft) modes whose eigenvalues lie on the real axis. Evidently, a scenario in which different modes close the gap indicates that the system experiences a series of different phase transitions where the dissipative gap closes in qualitatively different ways. The sequence of phase transitions as a function of $\tilde{\mathcal{E}}_1/\gamma_b$ between steady state and different time crystalline phases is highlighted in Fig. \ref{fig:ED_analysis}(a) using the colored background. Different phases were identified by the relative magnitude of the gap, the eigenvalues for the modes characterizing the time crystal phase [enclosed by arrows in Fig.~\ref{fig:ED_analysis}(c),(d)] and the period doubled modes [enclosed by rectangles in Fig.~\ref{fig:ED_analysis}(d)]. At a value of $\tilde{\mathcal{E}}_1/\gamma_b \approx 16$, the time crystal mode associated with the period doubled mode is practically as dominant as the usual time crystal mode. The edges of the coloring in Fig.~\ref{fig:ED_analysis} are blurred, as the exact behavior of the gap for larger values of $N$ and $\mathcal{E}_1$ are currently outside of computational capabilities. A full comparison of the periods of oscillatory quantum and semiclassical solutions is presented in Fig.~\ref{fig:ED_analysis}(e) (see also supplemental materials), where we also present the periods of the time crystalline phase obtained through Monte Carlo simulations~\cite{Qutip2} with the aim of investigating larger values of N.

To summarize, we compare semiclassical and quantum approaches: On Fig.~\ref{fig:sc_analysis} we observe the appearance of limit cycles for driving amplitudes between the points $\text{H}_1$ and $\text{H}_2$ which indicates a broken continuous time translation symmetry of the set of equations \eqref{CS_res_eqs_autonomous_vec} within this interval of $\mathcal{E}_1$. This is manifested by the discrete peak structure on Fig.~\ref{fig:sc_analysis} (d,e). In the quantum case we observe nearly non-decaying (almost zero real part of the Liouville eigenvalues) oscillating coherences at corresponding frequencies, see Fig.~\ref{fig:ED_analysis} (c,d). Semiclassical peaks on Fig.~\ref{fig:sc_analysis} (d,e) correspond to the points inside the thick red arrows and ovals on Fig.~\ref{fig:ED_analysis} (c,d) respectively.

\textit{Conclusions and discussions.}
In this work we demonstrated that a system of two coupled bosonic modes in a dissipative cavity exhibits rich behavior related to time crystalline phases. Based on the semiclassical approach, we have identified a parameter range in which a time crystalline phase emerges in the form of usual limit cycles or limit cycles featuring a doubled-loop structure associated with period doubling. Results of computations in quantum regime in the identified parameter range qualitatively agree with the global picture sketched by the semiclassical approach: A series of phase transitions is observed where oscillating coherences and period doubling modes emerge. These transitions are accompanied by the closure of the Liouvillean gap in the thermodynamic limit. Computational limitations do not allow us to probe the system at sufficiently large excitation number $N$ to make more precise, quantitative predictions on the phase transitions of the model discussed in this work. At this stage, experimental investigations, like in \cite{Zhang_2018,Xue_2019,Tikan_2021,Englebert_2021} are the natural next step for a detailed investigation of the predicted non-equilibrium phase transitions.

\textit{Acknowledgements.}
We would like to express our gratitude to the group of Dr. Philippe Corboz at the University of Amsterdam for allowing us to use their high capacity workstations to perform long-running computations on the quantum trajectories.
L.R.B., M.S.B., D.V.K., and A.K.F. thank the support by the Russian Science Foundation Grant No. 20-42-05002 (exact algebraic solution and semiclassical analysis) and the Russian Roadmap on Quantum Computing (exact diagonalization calculations). The work by V.G. is part of the DeltaITP consortium, a program of the Netherlands Organization for Scientific Research (NWO) funded by the Dutch Ministry of Education, Culture and Science (OCW).
Finally, this research was also supported by computational resources of HPC facilities at HSE University~\cite{HSEcomp}.

\vskip 2cm
\bibliography{Bibliography}

\begin{thebibliography}{60}%
\makeatletter
\providecommand \@ifxundefined [1]{%
 \@ifx{#1\undefined}
}%
\providecommand \@ifnum [1]{%
 \ifnum #1\expandafter \@firstoftwo
 \else \expandafter \@secondoftwo
 \fi
}%
\providecommand \@ifx [1]{%
 \ifx #1\expandafter \@firstoftwo
 \else \expandafter \@secondoftwo
 \fi
}%
\providecommand \natexlab [1]{#1}%
\providecommand \enquote  [1]{``#1''}%
\providecommand \bibnamefont  [1]{#1}%
\providecommand \bibfnamefont [1]{#1}%
\providecommand \citenamefont [1]{#1}%
\providecommand \href@noop [0]{\@secondoftwo}%
\providecommand \href [0]{\begingroup \@sanitize@url \@href}%
\providecommand \@href[1]{\@@startlink{#1}\@@href}%
\providecommand \@@href[1]{\endgroup#1\@@endlink}%
\providecommand \@sanitize@url [0]{\catcode `\\12\catcode `\$12\catcode
  `\&12\catcode `\#12\catcode `\^12\catcode `\_12\catcode `\%12\relax}%
\providecommand \@@startlink[1]{}%
\providecommand \@@endlink[0]{}%
\providecommand \url  [0]{\begingroup\@sanitize@url \@url }%
\providecommand \@url [1]{\endgroup\@href {#1}{\urlprefix }}%
\providecommand \urlprefix  [0]{URL }%
\providecommand \Eprint [0]{\href }%
\providecommand \doibase [0]{http://dx.doi.org/}%
\providecommand \selectlanguage [0]{\@gobble}%
\providecommand \bibinfo  [0]{\@secondoftwo}%
\providecommand \bibfield  [0]{\@secondoftwo}%
\providecommand \translation [1]{[#1]}%
\providecommand \BibitemOpen [0]{}%
\providecommand \bibitemStop [0]{}%
\providecommand \bibitemNoStop [0]{.\EOS\space}%
\providecommand \EOS [0]{\spacefactor3000\relax}%
\providecommand \BibitemShut  [1]{\csname bibitem#1\endcsname}%
\let\auto@bib@innerbib\@empty
\bibitem [{\citenamefont {Imoto}\ \emph {et~al.}(1985)\citenamefont {Imoto},
  \citenamefont {Haus},\ and\ \citenamefont {Yamamoto}}]{Imoto1985}%
  \BibitemOpen
  \bibfield  {author} {\bibinfo {author} {\bibfnamefont {N.}~\bibnamefont
  {Imoto}}, \bibinfo {author} {\bibfnamefont {H.~A.}\ \bibnamefont {Haus}}, \
  and\ \bibinfo {author} {\bibfnamefont {Y.}~\bibnamefont {Yamamoto}},\ }\href
  {\doibase 10.1103/PhysRevA.32.2287} {\bibfield  {journal} {\bibinfo
  {journal} {Phys. Rev. A}\ }\textbf {\bibinfo {volume} {32}},\ \bibinfo
  {pages} {2287} (\bibinfo {year} {1985})}\BibitemShut {NoStop}%
\bibitem [{\citenamefont {Harris}\ \emph {et~al.}(1990)\citenamefont {Harris},
  \citenamefont {Field},\ and\ \citenamefont {Imamo\ifmmode~\breve{g}\else
  \u{g}\fi{}lu}}]{Harris1990}%
  \BibitemOpen
  \bibfield  {author} {\bibinfo {author} {\bibfnamefont {S.~E.}\ \bibnamefont
  {Harris}}, \bibinfo {author} {\bibfnamefont {J.~E.}\ \bibnamefont {Field}}, \
  and\ \bibinfo {author} {\bibfnamefont {A.}~\bibnamefont
  {Imamo\ifmmode~\breve{g}\else \u{g}\fi{}lu}},\ }\href {\doibase
  10.1103/PhysRevLett.64.1107} {\bibfield  {journal} {\bibinfo  {journal}
  {Phys. Rev. Lett.}\ }\textbf {\bibinfo {volume} {64}},\ \bibinfo {pages}
  {1107} (\bibinfo {year} {1990})}\BibitemShut {NoStop}%
\bibitem [{\citenamefont {Turchette}\ \emph {et~al.}(1995)\citenamefont
  {Turchette}, \citenamefont {Hood}, \citenamefont {Lange}, \citenamefont
  {Mabuchi},\ and\ \citenamefont {Kimble}}]{Turchette1995}%
  \BibitemOpen
  \bibfield  {author} {\bibinfo {author} {\bibfnamefont {Q.~A.}\ \bibnamefont
  {Turchette}}, \bibinfo {author} {\bibfnamefont {C.~J.}\ \bibnamefont {Hood}},
  \bibinfo {author} {\bibfnamefont {W.}~\bibnamefont {Lange}}, \bibinfo
  {author} {\bibfnamefont {H.}~\bibnamefont {Mabuchi}}, \ and\ \bibinfo
  {author} {\bibfnamefont {H.~J.}\ \bibnamefont {Kimble}},\ }\href {\doibase
  10.1103/PhysRevLett.75.4710} {\bibfield  {journal} {\bibinfo  {journal}
  {Phys. Rev. Lett.}\ }\textbf {\bibinfo {volume} {75}},\ \bibinfo {pages}
  {4710} (\bibinfo {year} {1995})}\BibitemShut {NoStop}%
\bibitem [{\citenamefont {Kumar}\ and\ \citenamefont
  {DiVincenzo}(2010)}]{Kumar2010}%
  \BibitemOpen
  \bibfield  {author} {\bibinfo {author} {\bibfnamefont {S.}~\bibnamefont
  {Kumar}}\ and\ \bibinfo {author} {\bibfnamefont {D.~P.}\ \bibnamefont
  {DiVincenzo}},\ }\href {\doibase 10.1103/PhysRevB.82.014512} {\bibfield
  {journal} {\bibinfo  {journal} {Phys. Rev. B}\ }\textbf {\bibinfo {volume}
  {82}},\ \bibinfo {pages} {014512} (\bibinfo {year} {2010})}\BibitemShut
  {NoStop}%
\bibitem [{\citenamefont {Chang}\ \emph {et~al.}(2014)\citenamefont {Chang},
  \citenamefont {Vuleti{\'{c}}},\ and\ \citenamefont {Lukin}}]{Chang2014}%
  \BibitemOpen
  \bibfield  {author} {\bibinfo {author} {\bibfnamefont {D.~E.}\ \bibnamefont
  {Chang}}, \bibinfo {author} {\bibfnamefont {V.}~\bibnamefont
  {Vuleti{\'{c}}}}, \ and\ \bibinfo {author} {\bibfnamefont {M.~D.}\
  \bibnamefont {Lukin}},\ }\href {\doibase 10.1038/nphoton.2014.192} {\bibfield
   {journal} {\bibinfo  {journal} {Nat. Photonics}\ }\textbf {\bibinfo {volume}
  {8}},\ \bibinfo {pages} {685} (\bibinfo {year} {2014})}\BibitemShut {NoStop}%
\bibitem [{\citenamefont {England}\ \emph {et~al.}(2021)\citenamefont
  {England}, \citenamefont {Bouchard}, \citenamefont {Fenwick}, \citenamefont
  {Bonsma-Fisher}, \citenamefont {Zhang}, \citenamefont {Bustard},\ and\
  \citenamefont {Sussman}}]{doi:10.1063/5.0065222}%
  \BibitemOpen
  \bibfield  {author} {\bibinfo {author} {\bibfnamefont {D.}~\bibnamefont
  {England}}, \bibinfo {author} {\bibfnamefont {F.}~\bibnamefont {Bouchard}},
  \bibinfo {author} {\bibfnamefont {K.}~\bibnamefont {Fenwick}}, \bibinfo
  {author} {\bibfnamefont {K.}~\bibnamefont {Bonsma-Fisher}}, \bibinfo {author}
  {\bibfnamefont {Y.}~\bibnamefont {Zhang}}, \bibinfo {author} {\bibfnamefont
  {P.~J.}\ \bibnamefont {Bustard}}, \ and\ \bibinfo {author} {\bibfnamefont
  {B.~J.}\ \bibnamefont {Sussman}},\ }\href {\doibase 10.1063/5.0065222}
  {\bibfield  {journal} {\bibinfo  {journal} {Appl. Phys. Lett.}\ }\textbf
  {\bibinfo {volume} {119}},\ \bibinfo {pages} {160501} (\bibinfo {year}
  {2021})}\BibitemShut {NoStop}%
\bibitem [{\citenamefont {Zhang}\ \emph {et~al.}(2018)\citenamefont {Zhang},
  \citenamefont {Wang}, \citenamefont {Hu}, \citenamefont {Shams-Ansari},
  \citenamefont {Ren}, \citenamefont {Fan},\ and\ \citenamefont
  {Lon{\v{c}}ar}}]{Zhang_2018}%
  \BibitemOpen
  \bibfield  {author} {\bibinfo {author} {\bibfnamefont {M.}~\bibnamefont
  {Zhang}}, \bibinfo {author} {\bibfnamefont {C.}~\bibnamefont {Wang}},
  \bibinfo {author} {\bibfnamefont {Y.}~\bibnamefont {Hu}}, \bibinfo {author}
  {\bibfnamefont {A.}~\bibnamefont {Shams-Ansari}}, \bibinfo {author}
  {\bibfnamefont {T.}~\bibnamefont {Ren}}, \bibinfo {author} {\bibfnamefont
  {S.}~\bibnamefont {Fan}}, \ and\ \bibinfo {author} {\bibfnamefont
  {M.}~\bibnamefont {Lon{\v{c}}ar}},\ }\href {\doibase
  10.1038/s41566-018-0317-y} {\bibfield  {journal} {\bibinfo  {journal} {Nat.
  Photonics}\ }\textbf {\bibinfo {volume} {13}},\ \bibinfo {pages} {36}
  (\bibinfo {year} {2018})}\BibitemShut {NoStop}%
\bibitem [{\citenamefont {Xue}\ \emph {et~al.}(2019)\citenamefont {Xue},
  \citenamefont {Zheng},\ and\ \citenamefont {Zhou}}]{Xue_2019}%
  \BibitemOpen
  \bibfield  {author} {\bibinfo {author} {\bibfnamefont {X.}~\bibnamefont
  {Xue}}, \bibinfo {author} {\bibfnamefont {X.}~\bibnamefont {Zheng}}, \ and\
  \bibinfo {author} {\bibfnamefont {B.}~\bibnamefont {Zhou}},\ }\href {\doibase
  10.1038/s41566-019-0436-0} {\bibfield  {journal} {\bibinfo  {journal} {Nat.
  Photonics}\ }\textbf {\bibinfo {volume} {13}},\ \bibinfo {pages} {616}
  (\bibinfo {year} {2019})}\BibitemShut {NoStop}%
\bibitem [{\citenamefont {Xu}\ \emph {et~al.}(2019)\citenamefont {Xu},
  \citenamefont {Tan}, \citenamefont {Wu}, \citenamefont {Morandotti},
  \citenamefont {Mitchell},\ and\ \citenamefont {Moss}}]{kerr-rev}%
  \BibitemOpen
  \bibfield  {author} {\bibinfo {author} {\bibfnamefont {X.}~\bibnamefont
  {Xu}}, \bibinfo {author} {\bibfnamefont {M.}~\bibnamefont {Tan}}, \bibinfo
  {author} {\bibfnamefont {J.}~\bibnamefont {Wu}}, \bibinfo {author}
  {\bibfnamefont {R.}~\bibnamefont {Morandotti}}, \bibinfo {author}
  {\bibfnamefont {A.}~\bibnamefont {Mitchell}}, \ and\ \bibinfo {author}
  {\bibfnamefont {D.~J.}\ \bibnamefont {Moss}},\ }\href {\doibase
  10.1109/LPT.2019.2940497} {\bibfield  {journal} {\bibinfo  {journal} {IEEE
  Photon. Technol. Lett.}\ }\textbf {\bibinfo {volume} {31}},\ \bibinfo {pages}
  {1854} (\bibinfo {year} {2019})}\BibitemShut {NoStop}%
\bibitem [{\citenamefont {Tikan}\ \emph {et~al.}(2021)\citenamefont {Tikan},
  \citenamefont {Riemensberger}, \citenamefont {Komagata}, \citenamefont
  {Hönl}, \citenamefont {Churaev}, \citenamefont {Skehan}, \citenamefont
  {Guo}, \citenamefont {Wang}, \citenamefont {Liu}, \citenamefont {Seidler},\
  and\ \citenamefont {Kippenberg}}]{Tikan_2021}%
  \BibitemOpen
  \bibfield  {author} {\bibinfo {author} {\bibfnamefont {A.}~\bibnamefont
  {Tikan}}, \bibinfo {author} {\bibfnamefont {J.}~\bibnamefont
  {Riemensberger}}, \bibinfo {author} {\bibfnamefont {K.}~\bibnamefont
  {Komagata}}, \bibinfo {author} {\bibfnamefont {S.}~\bibnamefont {Hönl}},
  \bibinfo {author} {\bibfnamefont {M.}~\bibnamefont {Churaev}}, \bibinfo
  {author} {\bibfnamefont {C.}~\bibnamefont {Skehan}}, \bibinfo {author}
  {\bibfnamefont {H.}~\bibnamefont {Guo}}, \bibinfo {author} {\bibfnamefont
  {R.~N.}\ \bibnamefont {Wang}}, \bibinfo {author} {\bibfnamefont
  {J.}~\bibnamefont {Liu}}, \bibinfo {author} {\bibfnamefont {P.}~\bibnamefont
  {Seidler}}, \ and\ \bibinfo {author} {\bibfnamefont {T.~J.}\ \bibnamefont
  {Kippenberg}},\ }\href {\doibase 10.1038/s41567-020-01159-y} {\bibfield
  {journal} {\bibinfo  {journal} {Nat. Phys.}\ }\textbf {\bibinfo {volume}
  {17}},\ \bibinfo {pages} {604} (\bibinfo {year} {2021})}\BibitemShut
  {NoStop}%
\bibitem [{\citenamefont {Englebert}\ \emph {et~al.}(2021)\citenamefont
  {Englebert}, \citenamefont {Lucia}, \citenamefont {Parra-Rivas},
  \citenamefont {Arab{\'{\i}}}, \citenamefont {Sazio}, \citenamefont {Gorza},\
  and\ \citenamefont {Leo}}]{Englebert_2021}%
  \BibitemOpen
  \bibfield  {author} {\bibinfo {author} {\bibfnamefont {N.}~\bibnamefont
  {Englebert}}, \bibinfo {author} {\bibfnamefont {F.~D.}\ \bibnamefont
  {Lucia}}, \bibinfo {author} {\bibfnamefont {P.}~\bibnamefont {Parra-Rivas}},
  \bibinfo {author} {\bibfnamefont {C.~M.}\ \bibnamefont {Arab{\'{\i}}}},
  \bibinfo {author} {\bibfnamefont {P.-J.}\ \bibnamefont {Sazio}}, \bibinfo
  {author} {\bibfnamefont {S.-P.}\ \bibnamefont {Gorza}}, \ and\ \bibinfo
  {author} {\bibfnamefont {F.}~\bibnamefont {Leo}},\ }\href {\doibase
  10.1038/s41566-021-00858-z} {\bibfield  {journal} {\bibinfo  {journal} {Nat.
  Photonics}\ }\textbf {\bibinfo {volume} {15}},\ \bibinfo {pages} {857}
  (\bibinfo {year} {2021})}\BibitemShut {NoStop}%
\bibitem [{\citenamefont {Gottesman}\ \emph {et~al.}(2001)\citenamefont
  {Gottesman}, \citenamefont {Kitaev},\ and\ \citenamefont
  {Preskill}}]{GKP2001}%
  \BibitemOpen
  \bibfield  {author} {\bibinfo {author} {\bibfnamefont {D.}~\bibnamefont
  {Gottesman}}, \bibinfo {author} {\bibfnamefont {A.}~\bibnamefont {Kitaev}}, \
  and\ \bibinfo {author} {\bibfnamefont {J.}~\bibnamefont {Preskill}},\ }\href
  {\doibase 10.1103/PhysRevA.64.012310} {\bibfield  {journal} {\bibinfo
  {journal} {Phys. Rev. A}\ }\textbf {\bibinfo {volume} {64}},\ \bibinfo
  {pages} {012310} (\bibinfo {year} {2001})}\BibitemShut {NoStop}%
\bibitem [{\citenamefont {Grimm}\ \emph {et~al.}(2020)\citenamefont {Grimm},
  \citenamefont {Frattini}, \citenamefont {Puri}, \citenamefont {Mundhada},
  \citenamefont {Touzard}, \citenamefont {Mirrahimi}, \citenamefont {Girvin},
  \citenamefont {Shankar},\ and\ \citenamefont {Devoret}}]{Grimm2020}%
  \BibitemOpen
  \bibfield  {author} {\bibinfo {author} {\bibfnamefont {A.}~\bibnamefont
  {Grimm}}, \bibinfo {author} {\bibfnamefont {N.~E.}\ \bibnamefont {Frattini}},
  \bibinfo {author} {\bibfnamefont {S.}~\bibnamefont {Puri}}, \bibinfo {author}
  {\bibfnamefont {S.~O.}\ \bibnamefont {Mundhada}}, \bibinfo {author}
  {\bibfnamefont {S.}~\bibnamefont {Touzard}}, \bibinfo {author} {\bibfnamefont
  {M.}~\bibnamefont {Mirrahimi}}, \bibinfo {author} {\bibfnamefont {S.~M.}\
  \bibnamefont {Girvin}}, \bibinfo {author} {\bibfnamefont {S.}~\bibnamefont
  {Shankar}}, \ and\ \bibinfo {author} {\bibfnamefont {M.~H.}\ \bibnamefont
  {Devoret}},\ }\href {\doibase 10.1038/s41586-020-2587-z} {\bibfield
  {journal} {\bibinfo  {journal} {Nature}\ }\textbf {\bibinfo {volume} {584}},\
  \bibinfo {pages} {205} (\bibinfo {year} {2020})}\BibitemShut {NoStop}%
\bibitem [{\citenamefont {Yurke}\ and\ \citenamefont
  {Stoler}(1986)}]{Yurke1986}%
  \BibitemOpen
  \bibfield  {author} {\bibinfo {author} {\bibfnamefont {B.}~\bibnamefont
  {Yurke}}\ and\ \bibinfo {author} {\bibfnamefont {D.}~\bibnamefont {Stoler}},\
  }\href {\doibase 10.1103/PhysRevLett.57.13} {\bibfield  {journal} {\bibinfo
  {journal} {Phys. Rev. Lett.}\ }\textbf {\bibinfo {volume} {57}},\ \bibinfo
  {pages} {13} (\bibinfo {year} {1986})}\BibitemShut {NoStop}%
\bibitem [{\citenamefont {Kirchmair}\ \emph {et~al.}(2013)\citenamefont
  {Kirchmair}, \citenamefont {Vlastakis}, \citenamefont {Leghtas},
  \citenamefont {Nigg}, \citenamefont {Paik}, \citenamefont {Ginossar},
  \citenamefont {Mirrahimi}, \citenamefont {Frunzio}, \citenamefont {Girvin},\
  and\ \citenamefont {Schoelkopf}}]{Kirchmair2013}%
  \BibitemOpen
  \bibfield  {author} {\bibinfo {author} {\bibfnamefont {G.}~\bibnamefont
  {Kirchmair}}, \bibinfo {author} {\bibfnamefont {B.}~\bibnamefont
  {Vlastakis}}, \bibinfo {author} {\bibfnamefont {Z.}~\bibnamefont {Leghtas}},
  \bibinfo {author} {\bibfnamefont {S.~E.}\ \bibnamefont {Nigg}}, \bibinfo
  {author} {\bibfnamefont {H.}~\bibnamefont {Paik}}, \bibinfo {author}
  {\bibfnamefont {E.}~\bibnamefont {Ginossar}}, \bibinfo {author}
  {\bibfnamefont {M.}~\bibnamefont {Mirrahimi}}, \bibinfo {author}
  {\bibfnamefont {L.}~\bibnamefont {Frunzio}}, \bibinfo {author} {\bibfnamefont
  {S.~M.}\ \bibnamefont {Girvin}}, \ and\ \bibinfo {author} {\bibfnamefont
  {R.~J.}\ \bibnamefont {Schoelkopf}},\ }\href {\doibase 10.1038/nature11902}
  {\bibfield  {journal} {\bibinfo  {journal} {Nature}\ }\textbf {\bibinfo
  {volume} {495}},\ \bibinfo {pages} {205} (\bibinfo {year}
  {2013})}\BibitemShut {NoStop}%
\bibitem [{\citenamefont {Alaeian}\ \emph {et~al.}(2021)\citenamefont
  {Alaeian}, \citenamefont {Giedke}, \citenamefont {Carusotto}, \citenamefont
  {L\"ow},\ and\ \citenamefont {Pfau}}]{Alaeian2021-1}%
  \BibitemOpen
  \bibfield  {author} {\bibinfo {author} {\bibfnamefont {H.}~\bibnamefont
  {Alaeian}}, \bibinfo {author} {\bibfnamefont {G.}~\bibnamefont {Giedke}},
  \bibinfo {author} {\bibfnamefont {I.}~\bibnamefont {Carusotto}}, \bibinfo
  {author} {\bibfnamefont {R.}~\bibnamefont {L\"ow}}, \ and\ \bibinfo {author}
  {\bibfnamefont {T.}~\bibnamefont {Pfau}},\ }\href {\doibase
  10.1103/PhysRevA.103.013712} {\bibfield  {journal} {\bibinfo  {journal}
  {Phys. Rev. A}\ }\textbf {\bibinfo {volume} {103}},\ \bibinfo {pages}
  {013712} (\bibinfo {year} {2021})}\BibitemShut {NoStop}%
\bibitem [{\citenamefont {Alaeian}\ \emph {et~al.}()\citenamefont {Alaeian},
  \citenamefont {Soriente}, \citenamefont {Najafi},\ and\ \citenamefont
  {Yelin}}]{Alaeian2021-2}%
  \BibitemOpen
  \bibfield  {author} {\bibinfo {author} {\bibfnamefont {H.}~\bibnamefont
  {Alaeian}}, \bibinfo {author} {\bibfnamefont {M.}~\bibnamefont {Soriente}},
  \bibinfo {author} {\bibfnamefont {K.}~\bibnamefont {Najafi}}, \ and\ \bibinfo
  {author} {\bibfnamefont {S.~F.}\ \bibnamefont {Yelin}},\ }\href
  {https://arxiv.org/abs/2106.04045} {\bibinfo  {journal} {arXiv:2106.04045}\
  }\BibitemShut {NoStop}%
\bibitem [{\citenamefont {Lled\'o}\ \emph {et~al.}(2019)\citenamefont
  {Lled\'o}, \citenamefont {Mavrogordatos},\ and\ \citenamefont
  {Szyma\ifmmode~\acute{n}\else \'{n}\fi{}ska}}]{Cristobal2019}%
  \BibitemOpen
\bibfield  {journal} {  }\bibfield  {author} {\bibinfo {author} {\bibfnamefont
  {C.}~\bibnamefont {Lled\'o}}, \bibinfo {author} {\bibfnamefont {T.~K.}\
  \bibnamefont {Mavrogordatos}}, \ and\ \bibinfo {author} {\bibfnamefont
  {M.~H.}\ \bibnamefont {Szyma\ifmmode~\acute{n}\else \'{n}\fi{}ska}},\ }\href
  {\doibase 10.1103/PhysRevB.100.054303} {\bibfield  {journal} {\bibinfo
  {journal} {Phys. Rev. B}\ }\textbf {\bibinfo {volume} {100}},\ \bibinfo
  {pages} {054303} (\bibinfo {year} {2019})}\BibitemShut {NoStop}%
\bibitem [{\citenamefont {Casteels}\ \emph {et~al.}(2017)\citenamefont
  {Casteels}, \citenamefont {Fazio},\ and\ \citenamefont {Ciuti}}]{Ciuti2017}%
  \BibitemOpen
  \bibfield  {author} {\bibinfo {author} {\bibfnamefont {W.}~\bibnamefont
  {Casteels}}, \bibinfo {author} {\bibfnamefont {R.}~\bibnamefont {Fazio}}, \
  and\ \bibinfo {author} {\bibfnamefont {C.}~\bibnamefont {Ciuti}},\ }\href
  {\doibase 10.1103/PhysRevA.95.012128} {\bibfield  {journal} {\bibinfo
  {journal} {Phys. Rev. A}\ }\textbf {\bibinfo {volume} {95}},\ \bibinfo
  {pages} {012128} (\bibinfo {year} {2017})}\BibitemShut {NoStop}%
\bibitem [{\citenamefont {Seibold}\ \emph {et~al.}(2020)\citenamefont
  {Seibold}, \citenamefont {Rota},\ and\ \citenamefont {Savona}}]{Seibold2020}%
  \BibitemOpen
  \bibfield  {author} {\bibinfo {author} {\bibfnamefont {K.}~\bibnamefont
  {Seibold}}, \bibinfo {author} {\bibfnamefont {R.}~\bibnamefont {Rota}}, \
  and\ \bibinfo {author} {\bibfnamefont {V.}~\bibnamefont {Savona}},\ }\href
  {\doibase 10.1103/PhysRevA.101.033839} {\bibfield  {journal} {\bibinfo
  {journal} {Phys. Rev. A}\ }\textbf {\bibinfo {volume} {101}},\ \bibinfo
  {pages} {033839} (\bibinfo {year} {2020})}\BibitemShut {NoStop}%
\bibitem [{\citenamefont {Muniz}\ \emph {et~al.}(2020)\citenamefont {Muniz},
  \citenamefont {Barberena}, \citenamefont {Lewis-Swan}, \citenamefont {Young},
  \citenamefont {Cline}, \citenamefont {Rey},\ and\ \citenamefont
  {Thompson}}]{Muniz2020}%
  \BibitemOpen
  \bibfield  {author} {\bibinfo {author} {\bibfnamefont {J.~A.}\ \bibnamefont
  {Muniz}}, \bibinfo {author} {\bibfnamefont {D.}~\bibnamefont {Barberena}},
  \bibinfo {author} {\bibfnamefont {R.~J.}\ \bibnamefont {Lewis-Swan}},
  \bibinfo {author} {\bibfnamefont {D.~J.}\ \bibnamefont {Young}}, \bibinfo
  {author} {\bibfnamefont {J.~R.~K.}\ \bibnamefont {Cline}}, \bibinfo {author}
  {\bibfnamefont {A.~M.}\ \bibnamefont {Rey}}, \ and\ \bibinfo {author}
  {\bibfnamefont {J.~K.}\ \bibnamefont {Thompson}},\ }\href {\doibase
  10.1038/s41586-020-2224-x} {\bibfield  {journal} {\bibinfo  {journal}
  {Nature}\ }\textbf {\bibinfo {volume} {580}},\ \bibinfo {pages} {602}
  (\bibinfo {year} {2020})}\BibitemShut {NoStop}%
\bibitem [{\citenamefont {Lled{\'{o}}}\ and\ \citenamefont
  {Szyma{\'{n}}ska}(2020)}]{Cristobal2020}%
  \BibitemOpen
  \bibfield  {author} {\bibinfo {author} {\bibfnamefont {C.}~\bibnamefont
  {Lled{\'{o}}}}\ and\ \bibinfo {author} {\bibfnamefont {M.~H.}\ \bibnamefont
  {Szyma{\'{n}}ska}},\ }\href {\doibase 10.1088/1367-2630/ab9ae3} {\bibfield
  {journal} {\bibinfo  {journal} {New J. Phys.}\ }\textbf {\bibinfo {volume}
  {22}},\ \bibinfo {pages} {075002} (\bibinfo {year} {2020})}\BibitemShut
  {NoStop}%
\bibitem [{\citenamefont {Roberts}\ and\ \citenamefont
  {Clerk}(2020)}]{Roberts2020}%
  \BibitemOpen
  \bibfield  {author} {\bibinfo {author} {\bibfnamefont {D.}~\bibnamefont
  {Roberts}}\ and\ \bibinfo {author} {\bibfnamefont {A.~A.}\ \bibnamefont
  {Clerk}},\ }\href {\doibase 10.1103/PhysRevX.10.021022} {\bibfield  {journal}
  {\bibinfo  {journal} {Phys. Rev. X}\ }\textbf {\bibinfo {volume} {10}},\
  \bibinfo {pages} {021022} (\bibinfo {year} {2020})}\BibitemShut {NoStop}%
\bibitem [{\citenamefont {Gong}\ \emph {et~al.}(2018)\citenamefont {Gong},
  \citenamefont {Hamazaki},\ and\ \citenamefont {Ueda}}]{Gong2018}%
  \BibitemOpen
  \bibfield  {author} {\bibinfo {author} {\bibfnamefont {Z.}~\bibnamefont
  {Gong}}, \bibinfo {author} {\bibfnamefont {R.}~\bibnamefont {Hamazaki}}, \
  and\ \bibinfo {author} {\bibfnamefont {M.}~\bibnamefont {Ueda}},\ }\href
  {\doibase 10.1103/PhysRevLett.120.040404} {\bibfield  {journal} {\bibinfo
  {journal} {Phys. Rev. Lett.}\ }\textbf {\bibinfo {volume} {120}},\ \bibinfo
  {pages} {040404} (\bibinfo {year} {2018})}\BibitemShut {NoStop}%
\bibitem [{\citenamefont {Ippoliti}\ \emph {et~al.}(2021)\citenamefont
  {Ippoliti}, \citenamefont {Kechedzhi}, \citenamefont {Moessner},
  \citenamefont {Sondhi},\ and\ \citenamefont {Khemani}}]{Ippoliti2021}%
  \BibitemOpen
  \bibfield  {author} {\bibinfo {author} {\bibfnamefont {M.}~\bibnamefont
  {Ippoliti}}, \bibinfo {author} {\bibfnamefont {K.}~\bibnamefont {Kechedzhi}},
  \bibinfo {author} {\bibfnamefont {R.}~\bibnamefont {Moessner}}, \bibinfo
  {author} {\bibfnamefont {S.}~\bibnamefont {Sondhi}}, \ and\ \bibinfo {author}
  {\bibfnamefont {V.}~\bibnamefont {Khemani}},\ }\href {\doibase
  10.1103/PRXQuantum.2.030346} {\bibfield  {journal} {\bibinfo  {journal} {PRX
  Quantum}\ }\textbf {\bibinfo {volume} {2}},\ \bibinfo {pages} {030346}
  (\bibinfo {year} {2021})}\BibitemShut {NoStop}%
\bibitem [{\citenamefont {Cosme}\ \emph {et~al.}(2019)\citenamefont {Cosme},
  \citenamefont {Skulte},\ and\ \citenamefont {Mathey}}]{Cosme2019}%
  \BibitemOpen
  \bibfield  {author} {\bibinfo {author} {\bibfnamefont {J.~G.}\ \bibnamefont
  {Cosme}}, \bibinfo {author} {\bibfnamefont {J.}~\bibnamefont {Skulte}}, \
  and\ \bibinfo {author} {\bibfnamefont {L.}~\bibnamefont {Mathey}},\ }\href
  {\doibase 10.1103/PhysRevA.100.053615} {\bibfield  {journal} {\bibinfo
  {journal} {Phys. Rev. A}\ }\textbf {\bibinfo {volume} {100}},\ \bibinfo
  {pages} {053615} (\bibinfo {year} {2019})}\BibitemShut {NoStop}%
\bibitem [{\citenamefont {Ke{\ss}ler}\ \emph {et~al.}(2020)\citenamefont
  {Ke{\ss}ler}, \citenamefont {Cosme}, \citenamefont {Georges}, \citenamefont
  {Mathey},\ and\ \citenamefont {Hemmerich}}]{Ke_ler_2020}%
  \BibitemOpen
  \bibfield  {author} {\bibinfo {author} {\bibfnamefont {H.}~\bibnamefont
  {Ke{\ss}ler}}, \bibinfo {author} {\bibfnamefont {J.~G.}\ \bibnamefont
  {Cosme}}, \bibinfo {author} {\bibfnamefont {C.}~\bibnamefont {Georges}},
  \bibinfo {author} {\bibfnamefont {L.}~\bibnamefont {Mathey}}, \ and\ \bibinfo
  {author} {\bibfnamefont {A.}~\bibnamefont {Hemmerich}},\ }\href {\doibase
  10.1088/1367-2630/ab9fc0} {\bibfield  {journal} {\bibinfo  {journal} {New J.
  Phys.}\ }\textbf {\bibinfo {volume} {22}},\ \bibinfo {pages} {085002}
  (\bibinfo {year} {2020})}\BibitemShut {NoStop}%
\bibitem [{\citenamefont {Alaeian}\ and\ \citenamefont
  {Buča}()}]{Alaeian2022}%
  \BibitemOpen
  \bibfield  {author} {\bibinfo {author} {\bibfnamefont {H.}~\bibnamefont
  {Alaeian}}\ and\ \bibinfo {author} {\bibfnamefont {B.}~\bibnamefont
  {Buča}},\ }\href {https://arxiv.org/abs/2202.09369} {\bibinfo  {journal}
  {arXiv:2202.09369}\ }\BibitemShut {NoStop}%
\bibitem [{\citenamefont {Bu\ifmmode~\check{c}\else \v{c}\fi{}a}\ and\
  \citenamefont {Jaksch}(2019)}]{Buca2019-2}%
  \BibitemOpen
\bibfield  {journal} {  }\bibfield  {author} {\bibinfo {author} {\bibfnamefont
  {B.}~\bibnamefont {Bu\ifmmode~\check{c}\else \v{c}\fi{}a}}\ and\ \bibinfo
  {author} {\bibfnamefont {D.}~\bibnamefont {Jaksch}},\ }\href {\doibase
  10.1103/PhysRevLett.123.260401} {\bibfield  {journal} {\bibinfo  {journal}
  {Phys. Rev. Lett.}\ }\textbf {\bibinfo {volume} {123}},\ \bibinfo {pages}
  {260401} (\bibinfo {year} {2019})}\BibitemShut {NoStop}%
\bibitem [{\citenamefont {Bu{\v{c}}a}\ \emph {et~al.}(2019)\citenamefont
  {Bu{\v{c}}a}, \citenamefont {Tindall},\ and\ \citenamefont
  {Jaksch}}]{Buca2019-1}%
  \BibitemOpen
  \bibfield  {author} {\bibinfo {author} {\bibfnamefont {B.}~\bibnamefont
  {Bu{\v{c}}a}}, \bibinfo {author} {\bibfnamefont {J.}~\bibnamefont {Tindall}},
  \ and\ \bibinfo {author} {\bibfnamefont {D.}~\bibnamefont {Jaksch}},\ }\href
  {https://doi.org/10.1038/s41467-019-09757-y} {\bibfield  {journal} {\bibinfo
  {journal} {Nat. Commun.}\ }\textbf {\bibinfo {volume} {10}} (\bibinfo {year}
  {2019})}\BibitemShut {NoStop}%
\bibitem [{\citenamefont {Kelly}\ \emph {et~al.}(2021)\citenamefont {Kelly},
  \citenamefont {Timmermans}, \citenamefont {Marino},\ and\ \citenamefont
  {Tsai}}]{Kelly2021}%
  \BibitemOpen
  \bibfield  {author} {\bibinfo {author} {\bibfnamefont {S.~P.}\ \bibnamefont
  {Kelly}}, \bibinfo {author} {\bibfnamefont {E.}~\bibnamefont {Timmermans}},
  \bibinfo {author} {\bibfnamefont {J.}~\bibnamefont {Marino}}, \ and\ \bibinfo
  {author} {\bibfnamefont {S.-W.}\ \bibnamefont {Tsai}},\ }\href {\doibase
  10.21468/SciPostPhysCore.4.3.021} {\bibfield  {journal} {\bibinfo  {journal}
  {SciPost Phys. Core}\ }\textbf {\bibinfo {volume} {4}},\ \bibinfo {pages}
  {21} (\bibinfo {year} {2021})}\BibitemShut {NoStop}%
\bibitem [{\citenamefont {Yao}\ and\ \citenamefont {Nayak}(2018)}]{Yao2018}%
  \BibitemOpen
  \bibfield  {author} {\bibinfo {author} {\bibfnamefont {N.~Y.}\ \bibnamefont
  {Yao}}\ and\ \bibinfo {author} {\bibfnamefont {C.}~\bibnamefont {Nayak}},\
  }\href {\doibase 10.1063/pt.3.4020} {\bibfield  {journal} {\bibinfo
  {journal} {Phys. Today}\ }\textbf {\bibinfo {volume} {71}},\ \bibinfo {pages}
  {40} (\bibinfo {year} {2018})}\BibitemShut {NoStop}%
\bibitem [{\citenamefont {Zhang}\ \emph {et~al.}(2017)\citenamefont {Zhang},
  \citenamefont {Hess}, \citenamefont {Kyprianidis}, \citenamefont {Becker},
  \citenamefont {Lee}, \citenamefont {Smith}, \citenamefont {Pagano},
  \citenamefont {Potirniche}, \citenamefont {Potter}, \citenamefont
  {Vishwanath}, \citenamefont {Yao},\ and\ \citenamefont {Monroe}}]{Zhang2017}%
  \BibitemOpen
  \bibfield  {author} {\bibinfo {author} {\bibfnamefont {J.}~\bibnamefont
  {Zhang}}, \bibinfo {author} {\bibfnamefont {P.~W.}\ \bibnamefont {Hess}},
  \bibinfo {author} {\bibfnamefont {A.}~\bibnamefont {Kyprianidis}}, \bibinfo
  {author} {\bibfnamefont {P.}~\bibnamefont {Becker}}, \bibinfo {author}
  {\bibfnamefont {A.}~\bibnamefont {Lee}}, \bibinfo {author} {\bibfnamefont
  {J.}~\bibnamefont {Smith}}, \bibinfo {author} {\bibfnamefont
  {G.}~\bibnamefont {Pagano}}, \bibinfo {author} {\bibfnamefont {I.-D.}\
  \bibnamefont {Potirniche}}, \bibinfo {author} {\bibfnamefont {A.~C.}\
  \bibnamefont {Potter}}, \bibinfo {author} {\bibfnamefont {A.}~\bibnamefont
  {Vishwanath}}, \bibinfo {author} {\bibfnamefont {N.~Y.}\ \bibnamefont {Yao}},
  \ and\ \bibinfo {author} {\bibfnamefont {C.}~\bibnamefont {Monroe}},\ }\href
  {\doibase 10.1038/nature21413} {\bibfield  {journal} {\bibinfo  {journal}
  {Nature}\ }\textbf {\bibinfo {volume} {543}},\ \bibinfo {pages} {217}
  (\bibinfo {year} {2017})}\BibitemShut {NoStop}%
\bibitem [{\citenamefont {Choi}\ \emph {et~al.}(2017)\citenamefont {Choi},
  \citenamefont {Choi}, \citenamefont {Landig}, \citenamefont {Kucsko},
  \citenamefont {Zhou}, \citenamefont {Isoya}, \citenamefont {Jelezko},
  \citenamefont {Onoda}, \citenamefont {Sumiya}, \citenamefont {Khemani},
  \citenamefont {von Keyserlingk}, \citenamefont {Yao}, \citenamefont
  {Demler},\ and\ \citenamefont {Lukin}}]{Choi2017}%
  \BibitemOpen
  \bibfield  {author} {\bibinfo {author} {\bibfnamefont {S.}~\bibnamefont
  {Choi}}, \bibinfo {author} {\bibfnamefont {J.}~\bibnamefont {Choi}}, \bibinfo
  {author} {\bibfnamefont {R.}~\bibnamefont {Landig}}, \bibinfo {author}
  {\bibfnamefont {G.}~\bibnamefont {Kucsko}}, \bibinfo {author} {\bibfnamefont
  {H.}~\bibnamefont {Zhou}}, \bibinfo {author} {\bibfnamefont {J.}~\bibnamefont
  {Isoya}}, \bibinfo {author} {\bibfnamefont {F.}~\bibnamefont {Jelezko}},
  \bibinfo {author} {\bibfnamefont {S.}~\bibnamefont {Onoda}}, \bibinfo
  {author} {\bibfnamefont {H.}~\bibnamefont {Sumiya}}, \bibinfo {author}
  {\bibfnamefont {V.}~\bibnamefont {Khemani}}, \bibinfo {author} {\bibfnamefont
  {C.}~\bibnamefont {von Keyserlingk}}, \bibinfo {author} {\bibfnamefont
  {N.~Y.}\ \bibnamefont {Yao}}, \bibinfo {author} {\bibfnamefont
  {E.}~\bibnamefont {Demler}}, \ and\ \bibinfo {author} {\bibfnamefont {M.~D.}\
  \bibnamefont {Lukin}},\ }\href {\doibase 10.1038/nature21426} {\bibfield
  {journal} {\bibinfo  {journal} {Nature}\ }\textbf {\bibinfo {volume} {543}},\
  \bibinfo {pages} {221} (\bibinfo {year} {2017})}\BibitemShut {NoStop}%
\bibitem [{\citenamefont {Mi}\ and\ \citenamefont {{\em et
  al}}(2021)}]{Mi2022}%
  \BibitemOpen
  \bibfield  {author} {\bibinfo {author} {\bibfnamefont {X.}~\bibnamefont
  {Mi}}\ and\ \bibinfo {author} {\bibfnamefont {M.~I.}\ \bibnamefont {{\em et
  al}}},\ }\href {\doibase 10.1038/s41586-021-04257-w} {\bibfield  {journal}
  {\bibinfo  {journal} {Nature}\ }\textbf {\bibinfo {volume} {601}},\ \bibinfo
  {pages} {531} (\bibinfo {year} {2021})}\BibitemShut {NoStop}%
\bibitem [{\citenamefont {Ke\ss{}ler}\ \emph {et~al.}(2021)\citenamefont
  {Ke\ss{}ler}, \citenamefont {Kongkhambut}, \citenamefont {Georges},
  \citenamefont {Mathey}, \citenamefont {Cosme},\ and\ \citenamefont
  {Hemmerich}}]{Keler2021}%
  \BibitemOpen
  \bibfield  {author} {\bibinfo {author} {\bibfnamefont {H.}~\bibnamefont
  {Ke\ss{}ler}}, \bibinfo {author} {\bibfnamefont {P.}~\bibnamefont
  {Kongkhambut}}, \bibinfo {author} {\bibfnamefont {C.}~\bibnamefont
  {Georges}}, \bibinfo {author} {\bibfnamefont {L.}~\bibnamefont {Mathey}},
  \bibinfo {author} {\bibfnamefont {J.~G.}\ \bibnamefont {Cosme}}, \ and\
  \bibinfo {author} {\bibfnamefont {A.}~\bibnamefont {Hemmerich}},\ }\href
  {\doibase 10.1103/PhysRevLett.127.043602} {\bibfield  {journal} {\bibinfo
  {journal} {Phys. Rev. Lett.}\ }\textbf {\bibinfo {volume} {127}},\ \bibinfo
  {pages} {043602} (\bibinfo {year} {2021})}\BibitemShut {NoStop}%
\bibitem [{\citenamefont {Dogra}\ \emph {et~al.}(2019)\citenamefont {Dogra},
  \citenamefont {Landini}, \citenamefont {Kroeger}, \citenamefont {Hruby},
  \citenamefont {Donner},\ and\ \citenamefont {Esslinger}}]{Dogra2019}%
  \BibitemOpen
  \bibfield  {author} {\bibinfo {author} {\bibfnamefont {N.}~\bibnamefont
  {Dogra}}, \bibinfo {author} {\bibfnamefont {M.}~\bibnamefont {Landini}},
  \bibinfo {author} {\bibfnamefont {K.}~\bibnamefont {Kroeger}}, \bibinfo
  {author} {\bibfnamefont {L.}~\bibnamefont {Hruby}}, \bibinfo {author}
  {\bibfnamefont {T.}~\bibnamefont {Donner}}, \ and\ \bibinfo {author}
  {\bibfnamefont {T.}~\bibnamefont {Esslinger}},\ }\href {\doibase
  10.1126/science.aaw4465} {\bibfield  {journal} {\bibinfo  {journal}
  {Science}\ }\textbf {\bibinfo {volume} {366}},\ \bibinfo {pages} {1496}
  (\bibinfo {year} {2019})}\BibitemShut {NoStop}%
\bibitem [{\citenamefont {Wilczek}(2012)}]{Wilczek2012}%
  \BibitemOpen
  \bibfield  {author} {\bibinfo {author} {\bibfnamefont {F.}~\bibnamefont
  {Wilczek}},\ }\href {\doibase 10.1103/PhysRevLett.109.160401} {\bibfield
  {journal} {\bibinfo  {journal} {Phys. Rev. Lett.}\ }\textbf {\bibinfo
  {volume} {109}},\ \bibinfo {pages} {160401} (\bibinfo {year}
  {2012})}\BibitemShut {NoStop}%
\bibitem [{\citenamefont {Khemani}\ \emph {et~al.}()\citenamefont {Khemani},
  \citenamefont {Moessner},\ and\ \citenamefont {Sondhi}}]{khemani2019}%
  \BibitemOpen
  \bibfield  {author} {\bibinfo {author} {\bibfnamefont {V.}~\bibnamefont
  {Khemani}}, \bibinfo {author} {\bibfnamefont {R.}~\bibnamefont {Moessner}}, \
  and\ \bibinfo {author} {\bibfnamefont {S.~L.}\ \bibnamefont {Sondhi}},\
  }\href {https://arxiv.org/abs/1910.10745} {\ }\Eprint
  {http://arxiv.org/abs/1910.10745} {arXiv:1910.10745} \BibitemShut {NoStop}%
\bibitem [{\citenamefont {Sacha}\ and\ \citenamefont
  {Zakrzewski}(2017)}]{Sacha_2017}%
  \BibitemOpen
  \bibfield  {author} {\bibinfo {author} {\bibfnamefont {K.}~\bibnamefont
  {Sacha}}\ and\ \bibinfo {author} {\bibfnamefont {J.}~\bibnamefont
  {Zakrzewski}},\ }\href {\doibase 10.1088/1361-6633/aa8b38} {\bibfield
  {journal} {\bibinfo  {journal} {Rep. Prog. Phys.}\ }\textbf {\bibinfo
  {volume} {81}},\ \bibinfo {pages} {016401} (\bibinfo {year}
  {2017})}\BibitemShut {NoStop}%
\bibitem [{\citenamefont {Else}\ \emph {et~al.}(2020)\citenamefont {Else},
  \citenamefont {Monroe}, \citenamefont {Nayak},\ and\ \citenamefont
  {Yao}}]{Else2020}%
  \BibitemOpen
  \bibfield  {author} {\bibinfo {author} {\bibfnamefont {D.~V.}\ \bibnamefont
  {Else}}, \bibinfo {author} {\bibfnamefont {C.}~\bibnamefont {Monroe}},
  \bibinfo {author} {\bibfnamefont {C.}~\bibnamefont {Nayak}}, \ and\ \bibinfo
  {author} {\bibfnamefont {N.~Y.}\ \bibnamefont {Yao}},\ }\href {\doibase
  10.1146/annurev-conmatphys-031119-050658} {\bibfield  {journal} {\bibinfo
  {journal} {Annu. Rev. Condens. Matter Phys.}\ }\textbf {\bibinfo {volume}
  {11}},\ \bibinfo {pages} {467} (\bibinfo {year} {2020})}\BibitemShut
  {NoStop}%
\bibitem [{\citenamefont {Glendinning}(1994)}]{Glend}%
  \BibitemOpen
  \bibfield  {author} {\bibinfo {author} {\bibfnamefont {P.}~\bibnamefont
  {Glendinning}},\ }\href {\doibase 10.1017/CBO9780511626296} {\emph {\bibinfo
  {title} {Stability, Instability and Chaos: An Introduction to the Theory of
  Nonlinear Differential Equations}}},\ Cambridge Texts in Applied Mathematics\
  (\bibinfo  {publisher} {Cambridge University Press},\ \bibinfo {year}
  {1994})\BibitemShut {NoStop}%
\bibitem [{\citenamefont {Drummond}\ and\ \citenamefont
  {Walls}(1980)}]{Drummond1980}%
  \BibitemOpen
  \bibfield  {author} {\bibinfo {author} {\bibfnamefont {P.~D.}\ \bibnamefont
  {Drummond}}\ and\ \bibinfo {author} {\bibfnamefont {D.~F.}\ \bibnamefont
  {Walls}},\ }\href {\doibase 10.1088/0305-4470/13/2/034} {\bibfield  {journal}
  {\bibinfo  {journal} {J. Phys. A: Math. Gen.}\ }\textbf {\bibinfo {volume}
  {13}},\ \bibinfo {pages} {725} (\bibinfo {year} {1980})}\BibitemShut
  {NoStop}%
\bibitem [{\citenamefont {Bartolo}\ \emph {et~al.}(2016)\citenamefont
  {Bartolo}, \citenamefont {Minganti}, \citenamefont {Casteels},\ and\
  \citenamefont {Ciuti}}]{Bartolo2016}%
  \BibitemOpen
  \bibfield  {author} {\bibinfo {author} {\bibfnamefont {N.}~\bibnamefont
  {Bartolo}}, \bibinfo {author} {\bibfnamefont {F.}~\bibnamefont {Minganti}},
  \bibinfo {author} {\bibfnamefont {W.}~\bibnamefont {Casteels}}, \ and\
  \bibinfo {author} {\bibfnamefont {C.}~\bibnamefont {Ciuti}},\ }\href
  {\doibase 10.1103/PhysRevA.94.033841} {\bibfield  {journal} {\bibinfo
  {journal} {Phys. Rev. A}\ }\textbf {\bibinfo {volume} {94}},\ \bibinfo
  {pages} {033841} (\bibinfo {year} {2016})}\BibitemShut {NoStop}%
\bibitem [{\citenamefont {Gritsev}\ and\ \citenamefont
  {Polkovnikov}(2017)}]{Gritsev2017}%
  \BibitemOpen
  \bibfield  {author} {\bibinfo {author} {\bibfnamefont {V.}~\bibnamefont
  {Gritsev}}\ and\ \bibinfo {author} {\bibfnamefont {A.}~\bibnamefont
  {Polkovnikov}},\ }\href {\doibase 10.21468/SciPostPhys.2.3.021} {\bibfield
  {journal} {\bibinfo  {journal} {SciPost Phys.}\ }\textbf {\bibinfo {volume}
  {2}},\ \bibinfo {pages} {021} (\bibinfo {year} {2017})}\BibitemShut {NoStop}%
\bibitem [{\citenamefont {Ringel}\ and\ \citenamefont
  {Gritsev}(2013)}]{Ringel_2013}%
  \BibitemOpen
  \bibfield  {author} {\bibinfo {author} {\bibfnamefont {M.}~\bibnamefont
  {Ringel}}\ and\ \bibinfo {author} {\bibfnamefont {V.}~\bibnamefont
  {Gritsev}},\ }\href {\doibase 10.1103/PhysRevA.88.062105} {\bibfield
  {journal} {\bibinfo  {journal} {Phys. Rev. A}\ }\textbf {\bibinfo {volume}
  {88}},\ \bibinfo {pages} {062105} (\bibinfo {year} {2013})}\BibitemShut
  {NoStop}%
\bibitem [{\citenamefont {Bakker}\ \emph {et~al.}(2020)\citenamefont {Bakker},
  \citenamefont {Yashin}, \citenamefont {Kurlov}, \citenamefont {Fedorov},\
  and\ \citenamefont {Gritsev}}]{Bakker2020}%
  \BibitemOpen
  \bibfield  {author} {\bibinfo {author} {\bibfnamefont {L.~R.}\ \bibnamefont
  {Bakker}}, \bibinfo {author} {\bibfnamefont {V.~I.}\ \bibnamefont {Yashin}},
  \bibinfo {author} {\bibfnamefont {D.~V.}\ \bibnamefont {Kurlov}}, \bibinfo
  {author} {\bibfnamefont {A.~K.}\ \bibnamefont {Fedorov}}, \ and\ \bibinfo
  {author} {\bibfnamefont {V.}~\bibnamefont {Gritsev}},\ }\href {\doibase
  10.1103/PhysRevA.102.052220} {\bibfield  {journal} {\bibinfo  {journal}
  {Phys. Rev. A}\ }\textbf {\bibinfo {volume} {102}},\ \bibinfo {pages}
  {052220} (\bibinfo {year} {2020})}\BibitemShut {NoStop}%
\bibitem [{\citenamefont {Charzy{\'{n}}ski}\ and\ \citenamefont
  {Ku{\'{s}}}(2013)}]{Charzyski2013}%
  \BibitemOpen
  \bibfield  {author} {\bibinfo {author} {\bibfnamefont {S.}~\bibnamefont
  {Charzy{\'{n}}ski}}\ and\ \bibinfo {author} {\bibfnamefont {M.}~\bibnamefont
  {Ku{\'{s}}}},\ }\href {\doibase 10.1088/1751-8113/46/26/265208} {\bibfield
  {journal} {\bibinfo  {journal} {J. Phys. A: Math. Theor.}\ }\textbf {\bibinfo
  {volume} {46}},\ \bibinfo {pages} {265208} (\bibinfo {year}
  {2013})}\BibitemShut {NoStop}%
\bibitem [{\citenamefont {Wei}\ and\ \citenamefont {Norman}(1963)}]{Wei1963}%
  \BibitemOpen
  \bibfield  {author} {\bibinfo {author} {\bibfnamefont {J.}~\bibnamefont
  {Wei}}\ and\ \bibinfo {author} {\bibfnamefont {E.}~\bibnamefont {Norman}},\
  }\href {\doibase 10.1063/1.1703993} {\bibfield  {journal} {\bibinfo
  {journal} {J. Math. Phys.}\ }\textbf {\bibinfo {volume} {4}},\ \bibinfo
  {pages} {575} (\bibinfo {year} {1963})}\BibitemShut {NoStop}%
\bibitem [{\citenamefont {Wei}\ and\ \citenamefont {Norman}(1964)}]{Wei1964}%
  \BibitemOpen
  \bibfield  {author} {\bibinfo {author} {\bibfnamefont {J.}~\bibnamefont
  {Wei}}\ and\ \bibinfo {author} {\bibfnamefont {E.}~\bibnamefont {Norman}},\
  }\href {\doibase 10.1090/s0002-9939-1964-0160009-0} {\bibfield  {journal}
  {\bibinfo  {journal} {Proc. Am. Math. Soc.}\ }\textbf {\bibinfo {volume}
  {15}},\ \bibinfo {pages} {327} (\bibinfo {year} {1964})}\BibitemShut
  {NoStop}%
\bibitem [{\citenamefont {Scully}\ and\ \citenamefont
  {Zubairy}(1997)}]{scully_zubairy_1997}%
  \BibitemOpen
  \bibfield  {author} {\bibinfo {author} {\bibfnamefont {M.~O.}\ \bibnamefont
  {Scully}}\ and\ \bibinfo {author} {\bibfnamefont {M.~S.}\ \bibnamefont
  {Zubairy}},\ }\href {\doibase 10.1017/CBO9780511813993} {\emph {\bibinfo
  {title} {Quantum Optics}}}\ (\bibinfo  {publisher} {Cambridge University
  Press},\ \bibinfo {year} {1997})\BibitemShut {NoStop}%
\bibitem [{\citenamefont {Walls}\ and\ \citenamefont
  {Milburn}(2008)}]{Walls2006}%
  \BibitemOpen
  \bibinfo {editor} {\bibfnamefont {D.}~\bibnamefont {Walls}}\ and\ \bibinfo
  {editor} {\bibfnamefont {G.~J.}\ \bibnamefont {Milburn}},\ eds.,\ \href
  {\doibase 10.1007/978-3-540-28574-8} {\emph {\bibinfo {title} {Quantum
  Optics}}}\ (\bibinfo  {publisher} {Springer Berlin Heidelberg},\ \bibinfo
  {year} {2008})\BibitemShut {NoStop}%
\bibitem [{not()}]{note1}%
  \BibitemOpen
  \href@noop {} {}\bibinfo {note} {The presence of two-photon losses is not
  essential and can be omitted, at least at the semiclassical level. For
  completeness we keep it under consideration in this work.}\BibitemShut
  {Stop}%
\bibitem [{\citenamefont {Lindblad}(1976)}]{Lindblad1976}%
  \BibitemOpen
  \bibfield  {author} {\bibinfo {author} {\bibfnamefont {G.}~\bibnamefont
  {Lindblad}},\ }\href {\doibase 10.1007/bf01608499} {\bibfield  {journal}
  {\bibinfo  {journal} {Commun. Math. Phys.}\ }\textbf {\bibinfo {volume}
  {48}},\ \bibinfo {pages} {119} (\bibinfo {year} {1976})}\BibitemShut
  {NoStop}%
\bibitem [{\citenamefont {Dhooge}\ \emph {et~al.}(2008)\citenamefont {Dhooge},
  \citenamefont {Govaerts}, \citenamefont {Kuznetsov}, \citenamefont {Meijer},\
  and\ \citenamefont {Sautois}}]{matcont}%
  \BibitemOpen
  \bibfield  {author} {\bibinfo {author} {\bibfnamefont {A.}~\bibnamefont
  {Dhooge}}, \bibinfo {author} {\bibfnamefont {W.}~\bibnamefont {Govaerts}},
  \bibinfo {author} {\bibfnamefont {Y.~A.}\ \bibnamefont {Kuznetsov}}, \bibinfo
  {author} {\bibfnamefont {H.~G.}\ \bibnamefont {Meijer}}, \ and\ \bibinfo
  {author} {\bibfnamefont {B.}~\bibnamefont {Sautois}},\ }\href {\doibase
  10.1080/13873950701742754} {\bibfield  {journal} {\bibinfo  {journal} {Math.
  Comp. Model. Dyn. Sys.}\ }\textbf {\bibinfo {volume} {14}},\ \bibinfo {pages}
  {147} (\bibinfo {year} {2008})}\BibitemShut {NoStop}%
\bibitem [{\citenamefont {Krimer}\ and\ \citenamefont
  {Pletyukhov}(2019)}]{Krimer2019}%
  \BibitemOpen
  \bibfield  {author} {\bibinfo {author} {\bibfnamefont {D.~O.}\ \bibnamefont
  {Krimer}}\ and\ \bibinfo {author} {\bibfnamefont {M.}~\bibnamefont
  {Pletyukhov}},\ }\href {\doibase 10.1103/PhysRevLett.123.110604} {\bibfield
  {journal} {\bibinfo  {journal} {Phys. Rev. Lett.}\ }\textbf {\bibinfo
  {volume} {123}},\ \bibinfo {pages} {110604} (\bibinfo {year}
  {2019})}\BibitemShut {NoStop}%
\bibitem [{\citenamefont {Johansson}\ \emph {et~al.}(2013)\citenamefont
  {Johansson}, \citenamefont {Nation},\ and\ \citenamefont {Nori}}]{Qutip2}%
  \BibitemOpen
  \bibfield  {author} {\bibinfo {author} {\bibfnamefont {J.}~\bibnamefont
  {Johansson}}, \bibinfo {author} {\bibfnamefont {P.}~\bibnamefont {Nation}}, \
  and\ \bibinfo {author} {\bibfnamefont {F.}~\bibnamefont {Nori}},\ }\href
  {\doibase https://doi.org/10.1016/j.cpc.2012.11.019} {\bibfield  {journal}
  {\bibinfo  {journal} {Comput. Phys. Commun.}\ }\textbf {\bibinfo {volume}
  {184}},\ \bibinfo {pages} {1234} (\bibinfo {year} {2013})}\BibitemShut
  {NoStop}%
\bibitem [{\citenamefont {Guckenheimer}\ and\ \citenamefont
  {Holmes}(1983)}]{Guckenheimer1983}%
  \BibitemOpen
  \bibfield  {author} {\bibinfo {author} {\bibfnamefont {J.}~\bibnamefont
  {Guckenheimer}}\ and\ \bibinfo {author} {\bibfnamefont {P.}~\bibnamefont
  {Holmes}},\ }\href {\doibase 10.1007/978-1-4612-1140-2} {\emph {\bibinfo
  {title} {Nonlinear Oscillations, Dynamical Systems, and Bifurcations of
  Vector Fields}}}\ (\bibinfo  {publisher} {Springer New York},\ \bibinfo
  {year} {1983})\BibitemShut {NoStop}%
\bibitem [{\citenamefont {Carmichael}(2015)}]{Carmichael2015}%
  \BibitemOpen
  \bibfield  {author} {\bibinfo {author} {\bibfnamefont {H.~J.}\ \bibnamefont
  {Carmichael}},\ }\href {\doibase 10.1103/PhysRevX.5.031028} {\bibfield
  {journal} {\bibinfo  {journal} {Phys. Rev. X}\ }\textbf {\bibinfo {volume}
  {5}},\ \bibinfo {pages} {031028} (\bibinfo {year} {2015})}\BibitemShut
  {NoStop}%
\bibitem [{\citenamefont {Kostenetskiy}\ \emph {et~al.}(2021)\citenamefont
  {Kostenetskiy}, \citenamefont {Chulkevich},\ and\ \citenamefont
  {Kozyrev}}]{HSEcomp}%
  \BibitemOpen
  \bibfield  {author} {\bibinfo {author} {\bibfnamefont {P.~S.}\ \bibnamefont
  {Kostenetskiy}}, \bibinfo {author} {\bibfnamefont {R.~A.}\ \bibnamefont
  {Chulkevich}}, \ and\ \bibinfo {author} {\bibfnamefont {V.~I.}\ \bibnamefont
  {Kozyrev}},\ }\href {\doibase 10.1088/1742-6596/1740/1/012050} {\bibfield
  {journal} {\bibinfo  {journal} {J. Phys. Conf. Ser.}\ }\textbf {\bibinfo
  {volume} {1740}},\ \bibinfo {pages} {012050} (\bibinfo {year}
  {2021})}\BibitemShut {NoStop}%
\end{thebibliography}%


\begin{thebibliography}{99}%

\makeatletter
\providecommand \@ifxundefined [1]{%
 \@ifx{#1\undefined}
}%
\providecommand \@ifnum [1]{%
 \ifnum #1\expandafter \@firstoftwo
 \else \expandafter \@secondoftwo
 \fi
}%
\providecommand \@ifx [1]{%
 \ifx #1\expandafter \@firstoftwo
 \else \expandafter \@secondoftwo
 \fi
}%
\providecommand \natexlab [1]{#1}%
\providecommand \enquote  [1]{``#1''}%
\providecommand \bibnamefont  [1]{#1}%
\providecommand \bibfnamefont [1]{#1}%
\providecommand \citenamefont [1]{#1}%
\providecommand \href@noop [0]{\@secondoftwo}%
\providecommand \href [0]{\begingroup \@sanitize@url \@href}%
\providecommand \@href[1]{\@@startlink{#1}\@@href}%
\providecommand \@@href[1]{\endgroup#1\@@endlink}%
\providecommand \@sanitize@url [0]{\catcode `\\12\catcode `\$12\catcode
  `\&12\catcode `\#12\catcode `\^12\catcode `\_12\catcode `\%12\relax}%
\providecommand \@@startlink[1]{}%
\providecommand \@@endlink[0]{}%
\providecommand \url  [0]{\begingroup\@sanitize@url \@url }%
\providecommand \@url [1]{\endgroup\@href {#1}{\urlprefix }}%
\providecommand \urlprefix  [0]{URL }%
\providecommand \Eprint [0]{\href }%
\providecommand \doibase [0]{http://dx.doi.org/}%
\providecommand \selectlanguage [0]{\@gobble}%
\providecommand \bibinfo  [0]{\@secondoftwo}%
\providecommand \bibfield  [0]{\@secondoftwo}%
\providecommand \translation [1]{[#1]}%
\providecommand \BibitemOpen [0]{}%
\providecommand \bibitemStop [0]{}%
\providecommand \bibitemNoStop [0]{.\EOS\space}%
\providecommand \EOS [0]{\spacefactor3000\relax}%
\providecommand \BibitemShut  [1]{\csname bibitem#1\endcsname}%
\let\auto@bib@innerbib\@empty


\bibitem [{\citenamefont {Gritsev}\ and\ \citenamefont
  {Polkovnikov}(2017)}]{Gritsev2017-supp}%
  \BibitemOpen
  \bibfield  {author} {\bibinfo {author} {\bibfnamefont {V.}~\bibnamefont
  {Gritsev}}\ and\ \bibinfo {author} {\bibfnamefont {A.}~\bibnamefont
  {Polkovnikov}},\ }\href {\doibase 10.21468/SciPostPhys.2.3.021} {\bibfield
  {journal} {\bibinfo  {journal} {SciPost Phys.}\ }\textbf {\bibinfo {volume}
  {2}},\ \bibinfo {pages} {021} (\bibinfo {year} {2017})}\BibitemShut {NoStop}%
\bibitem [{\citenamefont {Ringel}\ and\ \citenamefont
  {Gritsev}(2013)}]{Ringel_2013-supp}%
  \BibitemOpen
  \bibfield  {author} {\bibinfo {author} {\bibfnamefont {M.}~\bibnamefont
  {Ringel}}\ and\ \bibinfo {author} {\bibfnamefont {V.}~\bibnamefont
  {Gritsev}},\ }\href {\doibase 10.1103/PhysRevA.88.062105} {\bibfield
  {journal} {\bibinfo  {journal} {Phys. Rev. A}\ }\textbf {\bibinfo {volume}
  {88}},\ \bibinfo {pages} {062105} (\bibinfo {year} {2013})}\BibitemShut
  {NoStop}%
  \bibitem [{\citenamefont {Bakker}\ \emph {et~al.}(2020)\citenamefont {Bakker},
  \citenamefont {Yashin}, \citenamefont {Kurlov}, \citenamefont {Fedorov},\
  and\ \citenamefont {Gritsev}}]{Bakker2020-supp}%
  \BibitemOpen
  \bibfield  {author} {\bibinfo {author} {\bibfnamefont {L.~R.}\ \bibnamefont
  {Bakker}}, \bibinfo {author} {\bibfnamefont {V.~I.}\ \bibnamefont {Yashin}},
  \bibinfo {author} {\bibfnamefont {D.~V.}\ \bibnamefont {Kurlov}}, \bibinfo
  {author} {\bibfnamefont {A.~K.}\ \bibnamefont {Fedorov}}, \ and\ \bibinfo
  {author} {\bibfnamefont {V.}~\bibnamefont {Gritsev}},\ }\href {\doibase
  10.1103/PhysRevA.102.052220} {\bibfield  {journal} {\bibinfo  {journal}
  {Phys. Rev. A}\ }\textbf {\bibinfo {volume} {102}},\ \bibinfo {pages}
  {052220} (\bibinfo {year} {2020})}\BibitemShut {NoStop}%
  \bibitem [{\citenamefont {Charzy{\'{n}}ski}\ and\ \citenamefont
  {Ku{\'{s}}}(2013)}]{Charzyski2013-supp}%
  \BibitemOpen
  \bibfield  {author} {\bibinfo {author} {\bibfnamefont {S.}~\bibnamefont
  {Charzy{\'{n}}ski}}\ and\ \bibinfo {author} {\bibfnamefont {M.}~\bibnamefont
  {Ku{\'{s}}}},\ }\href {\doibase 10.1088/1751-8113/46/26/265208} {\bibfield
  {journal} {\bibinfo  {journal} {J. Phys. A: Math. Theor.}\ }\textbf {\bibinfo
  {volume} {46}},\ \bibinfo {pages} {265208} (\bibinfo {year}
  {2013})}\BibitemShut {NoStop}%
  \bibitem [{\citenamefont {Wei}\ and\ \citenamefont {Norman}(1963)}]{Wei1963-supp}%
  \BibitemOpen
  \bibfield  {author} {\bibinfo {author} {\bibfnamefont {J.}~\bibnamefont
  {Wei}}\ and\ \bibinfo {author} {\bibfnamefont {E.}~\bibnamefont {Norman}},\
  }\href {\doibase 10.1063/1.1703993} {\bibfield  {journal} {\bibinfo
  {journal} {J. Math. Phys.}\ }\textbf {\bibinfo {volume} {4}},\ \bibinfo
  {pages} {575} (\bibinfo {year} {1963})}\BibitemShut {NoStop}%
\bibitem [{\citenamefont {Wei}\ and\ \citenamefont {Norman}(1964)}]{Wei1964-supp}%
  \BibitemOpen
  \bibfield  {author} {\bibinfo {author} {\bibfnamefont {J.}~\bibnamefont
  {Wei}}\ and\ \bibinfo {author} {\bibfnamefont {E.}~\bibnamefont {Norman}},\
  }\href {\doibase 10.1090/s0002-9939-1964-0160009-0} {\bibfield  {journal}
  {\bibinfo  {journal} {Proc. Am. Math. Soc.}\ }\textbf {\bibinfo {volume}
  {15}},\ \bibinfo {pages} {327} (\bibinfo {year} {1964})}\BibitemShut
  {NoStop}%
\bibitem [{\citenamefont {Scully}\ and\ \citenamefont
  {Zubairy}(1997)}]{scully_zubairy_1997-supp}%
  \BibitemOpen
  \bibfield  {author} {\bibinfo {author} {\bibfnamefont {M.~O.}\ \bibnamefont
  {Scully}}\ and\ \bibinfo {author} {\bibfnamefont {M.~S.}\ \bibnamefont
  {Zubairy}},\ }\href {\doibase 10.1017/CBO9780511813993} {\emph {\bibinfo
  {title} {Quantum Optics}}}\ (\bibinfo  {publisher} {Cambridge University
  Press},\ \bibinfo {year} {1997})\BibitemShut {NoStop}%
  \bibitem [{\citenamefont {Casteels}\ \emph {et~al.}(2017)\citenamefont
  {Casteels}, \citenamefont {Fazio},\ and\ \citenamefont {Ciuti}}]{Ciuti2017-supp}%
  \BibitemOpen
  \bibfield  {author} {\bibinfo {author} {\bibfnamefont {W.}~\bibnamefont
  {Casteels}}, \bibinfo {author} {\bibfnamefont {R.}~\bibnamefont {Fazio}}, \
  and\ \bibinfo {author} {\bibfnamefont {C.}~\bibnamefont {Ciuti}},\ }\href
  {\doibase 10.1103/PhysRevA.95.012128} {\bibfield  {journal} {\bibinfo
  {journal} {Phys. Rev. A}\ }\textbf {\bibinfo {volume} {95}},\ \bibinfo
  {pages} {012128} (\bibinfo {year} {2017})}\BibitemShut {NoStop}%
  \bibitem [{\citenamefont {Carmichael}(2015)}]{Carmichael2015-supp}%
  \BibitemOpen
  \bibfield  {author} {\bibinfo {author} {\bibfnamefont {H.~J.}\ \bibnamefont
  {Carmichael}},\ }\href {\doibase 10.1103/PhysRevX.5.031028} {\bibfield
  {journal} {\bibinfo  {journal} {Phys. Rev. X}\ }\textbf {\bibinfo {volume}
  {5}},\ \bibinfo {pages} {031028} (\bibinfo {year} {2015})}\BibitemShut
  {NoStop}%
  \bibitem [{\citenamefont {Pietracaprina}\ \emph {et~al.}(2018)\citenamefont
  {Pietracaprina}, \citenamefont {Macé}, \citenamefont {Luitz},\ and\
  \citenamefont {Alet}}]{Pietracaprina2018-supp}%
  \BibitemOpen
  \bibfield  {author} {\bibinfo {author} {\bibfnamefont {F.}~\bibnamefont
  {Pietracaprina}}, \bibinfo {author} {\bibfnamefont {N.}~\bibnamefont
  {Macé}}, \bibinfo {author} {\bibfnamefont {D.~J.}\ \bibnamefont {Luitz}}, \
  and\ \bibinfo {author} {\bibfnamefont {F.}~\bibnamefont {Alet}},\ }\href
  {\doibase 10.21468/SciPostPhys.5.5.045} {\bibfield  {journal} {\bibinfo
  {journal} {SciPost Phys.}\ }\textbf {\bibinfo {volume} {5}},\ \bibinfo
  {pages} {45} (\bibinfo {year} {2018})}\BibitemShut {NoStop}%
  \bibitem [{\citenamefont {Johansson}\ \emph {et~al.}(2013)\citenamefont
  {Johansson}, \citenamefont {Nation},\ and\ \citenamefont {Nori}}]{Qutip2-supp}%
  \BibitemOpen
  \bibfield  {author} {\bibinfo {author} {\bibfnamefont {J.}~\bibnamefont
  {Johansson}}, \bibinfo {author} {\bibfnamefont {P.}~\bibnamefont {Nation}}, \
  and\ \bibinfo {author} {\bibfnamefont {F.}~\bibnamefont {Nori}},\ }\href
  {\doibase https://doi.org/10.1016/j.cpc.2012.11.019} {\bibfield  {journal}
  {\bibinfo  {journal} {Comput. Phys. Commun.}\ }\textbf {\bibinfo {volume}
  {184}},\ \bibinfo {pages} {1234} (\bibinfo {year} {2013})}\BibitemShut
  {NoStop}%
\end{thebibliography}

\vspace{10cm}
\newpage
\mbox{}
\newpage
\onecolumngrid

\makeatletter
\setcounter{equation}{0}
\setcounter{figure}{0}
\setcounter{table}{0}
\setcounter{page}{1}
\renewcommand{\theequation}{S.\arabic{equation}}
\renewcommand{\thefigure}{S\arabic{figure}}
\renewcommand{\bibnumfmt}[1]{[S#1]}
\renewcommand{\citenumfont}[1]{S#1}

\section{Supplemental material}

\subsection{Lie algebraic solution of time dependent Lindblad equation}

Here, we describe in detail the Lie algebraic method used to solve the time dependent Lindblad equation \cite{Gritsev2017-supp,Ringel_2013-supp,Bakker2020-supp,Charzyski2013-supp,Wei1963-supp,Wei1964-supp,scully_zubairy_1997-supp}. 
Our starting point is to treat the $b$ mode semiclassically by replacing $\hat{b}$-operators with a c-number, $\hat{b} \rightarrow b$. This simplification allows us to solve the rest of the equation for the $a$-mode exactly using the Lie-algebraic approach since the remaining quantum operators form a closed Lie algebra. Using this approach we can keep track of any possible time dependencies exactly. Note that for time independent system parameters the semiclassical solution is in fact the exact solution to this system. For the driving protocol used in this work, time independence of the system parameters can be achieved through a rotating wave transformation. Here, however, we solve the system for any arbitrary choice of time dependent system parameters. For the $a$-mode from equation (2) in the main text we have
\be \label{supp_Lindblad_eq}
\begin{aligned}
	\dot \rho = - i \left[  \omega_a \hat a^{\dag}\hat a + g b^*(t) \hat a + g^* b(t) \hat a^{\dag}, \rho \right] +\frac{\gamma_a}{2} {\cal D}[\hat a] \rho\equiv {\cal L}\rho,
\end{aligned}
\ee
whereas for the $b$-mode in the semiclassical limit we find:
\be \label{supp_b_diff_eq}
	i \dot b = i \kappa_b\, b + g \langle \hat a \rangle + i K |b|^2 b + E_1(t)^*.
\ee
where $K = -\chi_b - i U$ and $\kappa_b = -\gamma_b/2 - i \omega_b$. Eqs.~(\ref{supp_Lindblad_eq}) and (\ref{supp_b_diff_eq}) form a coupled system. Together they provide a self-consistent solution to equation (2) in the main text in the absence of quantum fluctuations in the $b$-mode. Our quantum mechanical problem is therefore reduced to that of a simple harmonic oscillator with time dependent coherent drive and dissipation. In order to solve the system of Eqs~(\ref{supp_Lindblad_eq}) and (\ref{supp_b_diff_eq}) for any choice of the time dependent parameters, we can use the Lie-algebraic properties of the operators. The main idea behind this approach is that the time-evolution operator, being a time-ordered exponent, is an element of a Lie group, as long as the operators in the Liouvillian \eqref{supp_Lindblad_eq} form a closed Lie algebra. When this algebraic structure holds, one can make a solution ansatz of the following form:
\be \label{supp_rho_anzats}
	\rho(t) = \prod_j e^{c_j(t) O_j} \rho(0),
\ee
where $c_j(t)$ are time dependent functions that depend on the system parameters, and $\mathfrak{g}_j$ are the systems superoperators. Defining the commutator for two superoperators $O_{1}$ and~$O_{2}$ as 
\be
~[O_{1}, O_{2}]\rho=O_{1}(O_{2}\rho)-O_{2}(O_{1}\rho),
\ee
and identifying the superoperators in Eq.~\eqref{supp_Lindblad_eq} (i.e. the set of $O_j$'s in (\ref{supp_rho_anzats})) as
\be \label{supp_superops_def}
\begin{aligned}
&J\rho= a \,\rho  a^{\dag},   &&B_{L}\rho =a^{\dag}a\rho, \;\; &&B_{R} \rho=\rho a^{\dag}a,\\
&A_{L}\rho = a\rho,  &&A^{\dag}_{L}\rho=a^{\dag}\rho,\;\;  &&A^{\dag}_{R} \rho =\rho a, \qquad A_{R}\rho =\rho a^{\dag},
\end{aligned}
\ee
we find that these superoperators generate the following Lie algebra:
\be \label{supp_Lie_algebra}
\begin{aligned}
	&[A_{L}, A^{\dag}_{L}] = 1,\hspace{2cm} && [A_{R},A^{\dag}_{R}]=1\\
	&[B_{L}, A_{L}] = - A_{L},  &&[B_{R},A_{R}]=-A_{R},\\
	&[B_{L},A^{\dag}_{L}]=A^{\dag}_{L}, && [B_{R}, A^{\dag}_{R}] = A^{\dag}_{R}\\
	&[B_{L},J] =- J, &&[B_{R},J] =- J, \\
	&[J, A_{L}^{\dag}]=A_{R},&&[J, A^{\dag}_{R}] = A_{L} ,\\
\end{aligned}
\ee
with all other commutators being zero. In terms of these superoperators, Liouvillian \eqref{supp_Lindblad_eq} can be written as
\be \label{supp_Liouvillian_gens}
\begin{aligned}
	{\cal L} =& \, \kappa_a B_L + \kappa^*_a B_R + \gamma_a J + i g b^* \left( A_R^{\dag} - A_L \right) + i g^*b \left( A_R - A_L^{\dag} \right),
\end{aligned}
\ee
where we introduce the same parameters $\kappa_a = -i\omega_a -\gamma_a/2$ as in the main text.
Since the Liouvillian is generated by elements forming the closed Lie algebra, the solution to \eqref{supp_Lindblad_eq} is given by a product of {\it ordinary} (opposed to the time-ordered in the general case) time dependent exponentials of the form in eq. \eqref{supp_rho_anzats}
Obviously, the choice for this type of representation is not unique (e.g it could depend on the ordering of individual factors), however if for $\rho(0)$ we pick up a coherent initial state
\be\label{supp_rho0_CS}
\rho(0) = \ket{\alpha_0}\!\bra{\alpha_0},
\ee
then a natural choice for the ordering of the products would be the of the form
\be \label{supp_CS_Ansatz}
\begin{aligned}
	\rho(t) = \, & e^{f(t)}\,e^{h(t) J} \,e^{\beta(t) B_{L}} \, e^{\beta^*(t) B_{R}} e^{  \alpha(t) A_{L}^{\dag}}e^{-  \alpha^*(t) A_{L}} \,e^{\alpha^*(t) A^{\dag}_{R}}e^{-\alpha(t) A_{R}}\,\rho(0).
\end{aligned}
\ee
Here, the relationships between the different time dependent functions $c_{j}(t)$ can be derived from the hermiticity of the density matrix $\rho(t)$. Note the presence of the prefactor $\exp\left[f(t)\right]$. It appears because the Lie algebra in Eq.~(\ref{supp_Lie_algebra}) is spanned not only by the superoperators~(\ref{supp_superops_def}), but it also contains the identity operator, as can be seen from the commutation relations. The Ansatz in Eq.~(\ref{supp_CS_Ansatz}) is one of the most natural ones because the action of the group elements on the initial state (\ref{supp_rho0_CS}) is very simple. Differentiating $\rho(t)$ in Eq.~(\ref{supp_CS_Ansatz}) and using various adjoint actions (in order to commute all exponentials to the right), such as 
\be
\begin{array}{lll}
e^{g J}\,B_L\,e^{-g J} \;=\; B_L+g J,\qquad\qquad & e^{g J}\,B_R\,e^{-g J} \;=\; B_R+g J,\qquad\qquad &e^{g J}\,A_R^{\dag}\,e^{-g J} \;=\; A_R^{\dag} + g A_L,\\
e^{g J}\,A_L^{\dag}\,e^{-g J} \;=\; A_L^{\dag} + g A_R,& e^{g A_R}\,B_R\,e^{-g A_R} \;=\; B_R + g A_R, & e^{g A_R}\,A_R^{\dag}\,e^{-g A_R} \;=\; A_R^{\dag} + g,\\
e^{g A_R^{\dag}}\,J\,e^{-g A_R^{\dag}} \;=\; J-g A_L, & e^{g A_R^{\dag}}\,B_R\,e^{-g A_R^{\dag}} \;=\; B_R-g A_R^{\dag},& e^{g A_R^{\dag}}\,A_R\,e^{-g A_R} \;=\; A_R-g,\\
e^{g A_L}\,B_L\,e^{-g A_L} \;=\; B_L + g A_L,&
e^{g A_L}\,A_L^{\dag}\,e^{-g A_L} \;=\; A_L^{\dag} + g,
&e^{g A_L^{\dag}}\,J\,e^{-g A_L^{\dagger}} \;=\; J-g A_R,\\
e^{g A_L^{\dag}}\,B_L\,e^{-g  A_L^{\dagger}} \;=\; B_L-g A_L^{\dag},&e^{g A_L^{\dag}}\,A_L\,e^{-g  A_L^{\dagger}} \;=\; A_L-g, & e^{g B_R}\,J\, e^{-g B_R} \;=\; e^{-g} \, J,\\
e^{g B_R}\,A_R\,e^{-g B_R} \;=\; e^{-g} \, A_R, & e^{g B_R}\,A_R^{\dag}\,e^{-g B_R} \;=\; e^{g}\, A_R^{\dag},& e^{g B_L}\,J\,e^{-g B_L} \;=\; e^{-g}\,J,\\
e^{g B_L}\,A_L\,e^{-g B_L} \;=\; e^{-g}\,A_L,& e^{g B_L}\,A_L^{\dag}\,e^{-g B_L} \;=\; e^{g}\, A_L^{\dag},\\
\end{array}
\ee 
we get
\be
\begin{aligned}
	\dot\rho = & \,\Bigl\{ \dot\beta^* B_R + e^{\beta^*} \dot\alpha^* A_R^{\dag} + \left( h e^{\beta} - e^{-\beta^*} \right) \dot\alpha A_R + \dot\beta B_L + e^{\beta} \dot\alpha A_L^{\dag} \\
	&+ \left( h e^{\beta^*} - e^{-\beta} \right) \dot\alpha^* A_L + \left[\dot h + h\left( \dot\beta + \dot\beta^* \right) \right] J + \dot f + \partial_t |\alpha|^2 \Bigr\} \, \rho(t).
\end{aligned}
\ee
Matching the expression in the curly brackets with Liouvillian (\ref{supp_Liouvillian_gens}),
we obtain a system of differential equations for $f(t)$, $h(t)$, $\alpha(t)$, and $\beta(t)$:
\be
\label{supp_coefficients_diff_eqs}
\begin{aligned}
	\dot \beta &= \kappa_a,  &&\qquad\dot h - \gamma_a h - \gamma_a = 0, \\
	 \dot \alpha &= -i g^* b e^{-\beta},  &&\qquad \dot f + \partial_t |\alpha|^2 = 0,
\end{aligned}
\ee
with zero initial conditions. Note that the above equations are coupled with the semiclassical equation for the $b$-mode Eq.~\eqref{supp_b_diff_eq}. This particular system of differential equations, 
Eqs.~\eqref{supp_coefficients_diff_eqs}, can be solved in terms of quadratures as follows:
\be \label{supp_CS_DE_sol}
\begin{aligned}
	&\beta(t) = \int_0^t \,d\tau \, \kappa_a(\tau), &&\qquad \alpha(t) = - i \int_0^t \, d\tau\, e^{ -\beta(\tau) } g^*(\tau) b(\tau),\\
	&h(t) = e^{ -\left( \beta(t) + \beta^*(t) \right) } - 1 =e^{\int_0^t \,d\tau\,\gamma_a(\tau)} - 1,&&\qquad	f(t) = -|\alpha(t)|^2,\\
\end{aligned}
\ee
where $b(t)$ is a solution to Eq.~(\ref{supp_b_diff_eq}). Indeed, taking into account the definitions of $A_{L,R}\rho$ and $A_{L,R}^{\dag}\rho$, along with the expression for the displacement operator in terms of $a$ and $a^{\dag}$,
\be
	\ket{\alpha_0} = D\left(\alpha_0\right)\ket{0} = e^{-\left| \alpha_0 \right|^2/2} e^{\alpha_0 a^{\dag}}e^{-\alpha_0^* a}\ket{0},
\ee
where $\ket{\alpha_0}$ is a coherent state, the action of the last four exponents in Eq. (\ref{supp_CS_Ansatz}) can be written as
\be \label{group_action_1}
\begin{aligned}
	e^{|\alpha(t)|^2} D\left(\alpha(t)\right)\ket{\alpha_0}\!\bra{\alpha_0}D\left(-\alpha(t) \right) = e^{|\alpha(t)|^2} \ket{\alpha_0+\alpha(t)}\!\bra{\alpha_0+\alpha(t)}.
\end{aligned}
\ee
Here we took into account that $\ket{\alpha_0}\!\bra{\alpha_0} = D(\alpha_0)\ket{0}\!\bra{0}D(-\alpha_0)$ and used the following property of the displacement operators: $D(\alpha)D(\alpha_0) = \exp\{ \alpha \alpha_0^* - \alpha^* \alpha_0 \}D(\alpha_0+\alpha)$. 
Note that the exponential prefactor cancels with the factor of $e^{f(t)}$ in Eq. (\ref{supp_CS_Ansatz}).
It is now straightforward to calculate the action of the remaining exponentials in the first line of Eq. (\ref{supp_CS_Ansatz}) on the state $\ket{\alpha_0+\alpha(t)}\!\bra{\alpha_0+\alpha(t)}$. Using the fact that 
\be	
	e^{\beta^* a^{\dag} a}\ket{v} = e^{-|v|^2/2}\sum_{n=0}^{+\infty}\frac{\left( v\, e^{\beta^*} \right)^n}{\sqrt{n!}}\ket{n}
	= \exp\left\{ \frac{|v\, e^{\beta^*}|^2 - |v|^2}{2} \right\} \! \ket{v\, e^{\beta^*} },
\ee
where $\ket{v}$ is a coherent state, we get 
\be \label{group_action_2}
	e^{\beta^* B_{L}} \, e^{\beta B_{R}} \ket{\alpha_0+\alpha}\!\bra{\alpha_0+\alpha} 
	 = e^{ (e^{2\Re \beta}-1) \left| \alpha_0+\alpha\right|^2 } \bigl|(\alpha_0+\alpha) e^{\beta^*}\bigr> \bigl<(\alpha_0+\alpha) e^{\beta^*}\bigr|.
\ee
Finally, the action of $\exp(h J)$ is trivially found using the formal power series expansion of the exponent
\be
	e^{h J} \ket{v}\!\bra{v} = \sum_{n=0}^{+\infty} \,\frac{h^n}{n!} \, a^n \ket{v}\!\bra{v} \, \left(a^{\dag} \right)^n = e^{h |v|^2} \ket{v}\!\bra{v}.
\ee
In our case this yields
\be \label{group_action_3}
	e^{h J} \bigl|(\alpha_0+\alpha) e^{\beta^*}\bigr> \bigl<(\alpha_0+\alpha) e^{\beta^*}\bigr| 
	= \exp\left\{ h\, e^{2\text{Re}(\beta)} \left| \alpha_0+\alpha\right|^2 \right\} \bigl|(\alpha_0+\alpha) e^{\beta^*}\bigr> \bigl<(\alpha_0+\alpha) e^{\beta^*}\bigr|.
\ee
Then, taking into account that  $h~ \text{exp}\left[2\text{Re}(\beta)\right] = 1 - \text{exp}\left[2\text{Re}(\beta)\right]$, as follows from Eq. (\ref{supp_CS_DE_sol}), we combine Eqs. (\ref{group_action_1}), (\ref{group_action_2}), and (\ref{group_action_3}), and observe that all exponential prefactors cancel out. Thus, the density matrix from Eq. (\ref{supp_CS_Ansatz}) reduces to
\be \label{supp_CS_rho_t}
	\rho(t) = \bigl|\left(\alpha_0+\alpha(t)\right) e^{\beta(t)}\bigr> \bigl<\left(\alpha_0+\alpha(t)\right) e^{\beta(t)}\bigr|.
\ee
This exact solution allows us to compute the expectation value of $\hat{a}(t)$ as
\be
    \alpha(t) = \left<a(t)\right> = \Tr\left[\hat{a}\rho(t)\right]
\ee
thus effectively reducing our problem to solving the following set of coupled differential equations,
\be \label{supp_CS_res_eqs}
\begin{aligned}
	i \dot \alpha = &g^* e^{-\beta}  b,  \\
	i \dot b = &i \kappa_b\, b + g e^{\beta} \left( \alpha_0 + \alpha \right) + i K |b|^2 b + E_1(t)^*.
\end{aligned}
\ee
Our problem can further be reduced to an autonomous system of equations by assuming that $g$, $\kappa_{a}$, $\kappa_{b}$, and $U$ are time-independent, and the driving amplitude being an oscillatory function $E_1(t) = {\cal E}_1 e^{i \omega_1 t}$.
In this case, using an appropriate chance of variables, 
\be
\begin{aligned}
	y(t) &= e^{\beta(t) + i \omega_1 t   }\bigl( \alpha_0 + \alpha(t) \bigr),\\
	z(t) &= e^{i \omega_1 t } b(t),
\end{aligned}
\ee
one can reduce Eq.~(\ref{supp_CS_res_eqs}) to
\begin{equation}
    \begin{aligned}
        \dot{y} =& \tilde{\kappa}_a y - ig^* z,\\
        \dot{z} =& -igy + \tilde{\kappa}_b z + K|z|^2 z - i{\cal E}_1^*
    \end{aligned}
\end{equation}
which is used to find the results reported in equation (3) in the main text.

\subsection{Weak interaction limit}

The system we consider here is, technically speaking, a zero dimensional system, and as such  there is no obvious concept of taking a thermodynamic limit. However, Casteels et al. \cite{Ciuti2017-supp} made a inspirational argument based on the works of Carmicheal \cite{Carmichael2015-supp}: they compared the Fourier transforms of the Liouvillian \eqref{supp_Lindblad_eq} and a system of $N$ copies of dissipative coherently driven Bose-Hubbard chains, where $N$ is the number of cavities. This  resembles a thermodynamic limit. The suggested equivalence can readily be derived by substituting Fourier transformed bosonic operators into the Hamiltonian for the 1D Bose Hubbard chain. Note that the homogeneous drive corresponds only to the $k = 0$ mode in the expansion.
\begin{figure*}[t]
\includegraphics[width=\linewidth]{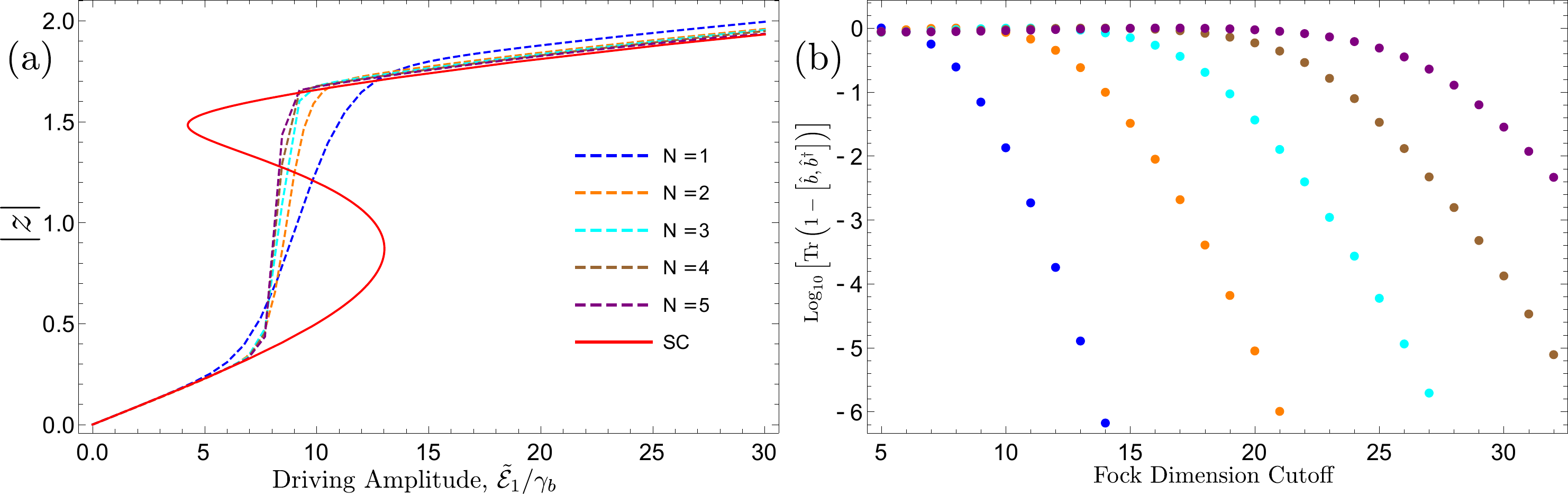}
 \caption{(a) Comparison of semi classical steady state solution as a function of $\tilde{\mathcal{E}}_1/\gamma_b$ (Red curve) and the numerically computed, exact solutions of the steady state using a truncated Fock space representation of the bosonic a and b modes (dashed curves).  The parameters are $g = 5$, $\arg {\cal E}_1 = 0$, $\gamma_a = \chi_b = 1$, $\Delta_a = U = 10$ and $\Delta_b = -20$ (in units of $\gamma_b$). Cutoff dimensionality for both modes is based on the results from Fig (b) and are 17, 17, 24, 28, 30 for the modes $N = 1\dots 5$ respectively. (b) Convergence of steady-state expectation value $\Tr\left(\hat{b}^\dagger\hat{b}\rho_{ss}\right)$ measured by evaluating $\Tr\left(1-[\hat{b}^\dagger,\hat{b}]\right)$ for different values of $N$. The closer this value is to zero, the better the convergence is. Fock space dimensionality cutoff is the same for both the $a$- and $b$-modes. The parameters are are the same as in Fig (a). The convergence is measured at the point $\tilde{\mathcal{E}}_1/\gamma_b = 30$ which requires the largest dimensionality of the Hilbert space to converge, since the steady state particle number increases as a function of $\tilde{\mathcal{E}}_1$.}\label{fig:supp_NumericalDiagonalization}
\end{figure*}
The only difference between the arguments made in \cite{Ciuti2017-supp} and this work is that we also added a two-mode dissipation channel. This results in rescaling the parameters $\mathcal{E}_1$, $U$ and the dissipation rate $\chi$. The scaling of $\chi$ turns out to be the same as that of $U$. Omitting the details of the trivial calculation we conclude that we need to rescale the system parameters as follows
\be
\mathcal{E}_1 = \tilde{\mathcal{E}}_1\sqrt{N},\qquad U = \frac{\tilde{U}}{N}, \qquad \chi = \frac{\tilde{\chi}}{N}.
\ee
Using this notion of thermodynamic limit we can perform numerical computations to probe the full quantum phase diagram of our system. Fig.~\ref{fig:supp_NumericalDiagonalization}(a) shows that the quantum mechanical description of the system converges to the semiclassical prediction as we  increase $N$, the steady state expectation value of the particle number operator. Note that in the quantum regime there is no bi-stability because of the single-valuedness of the wavefunction. Instead, we observe a sharp transition. The increase in the slope of the transition is expected to grow proportionally to $N$, as confirmed by this plot. The datapoints in Fig.~\ref{fig:supp_NumericalDiagonalization}(a) have been computed using a cutoff representation of the $a$- and $b$-mode operators in the Hilbert space. Convergence was carefully monitored to ensure that these results are trustworthy. An example of the convergence that was monitored is given in Fig~\ref{fig:supp_NumericalDiagonalization}(b). Here we were looking at the steady state expectation value of the operator $1-[\hat{b},\hat{b}^\dag]$ which should be zero for a high convergence. The datapoints in the figure are taken at $\tilde{{\cal E}}_1/\gamma_b = 3.0$, which requires the largest cutoff values to converge in the data set corresponding to this figure. We would like to note that the full quantum mechanical Liouvillian scales as $ \text{Rank}(\hat{a})^2\times \text{Rank}(\hat{b})^2$. By increasing the value of $N$, the steady state expectation values of $\Tr(\hat{a}^\dag \hat{a}\rho(t))$ and $\Tr(\hat{b}^\dag \hat{b}\rho(t))$ steadily increase and spread out, requiring larger cutoffs in the Hilbert space. Direct eigenvalue solver algorithms using LU decomposition (e.g. in the ARPACK library) scale rather expensively in their RAM consumption. For steady state calculations we can use a trick to increase the computational efficiency. We introduce a new matrix $O = L^\dagger L$,
where $L$ is the vectorized Liouvillian. The zero eigenvector $\vec{\rho}$ of $O$ is the steady-state solution of $L$. This can be shown by considering the relationship of the kernel of the new matrix $O\rho=L^\dag (L \vec{\rho}) = 0 \iff L\vec{\rho} \in \text{Ker}(L^\dag)$. Since isomorphisms leave kernels invariant, this implies that $L^\dag\vec{\rho} \in \text{Ker}(L) \implies L^\dag\vec{\rho}  = 0.$
Thus, the zero eigenvector of $O$ is the same as that of $L$ and $L^\dag$.
This hermitization of the density matrix $L$ allows us to compute eigenvalues and vectors for the Liouvillian $L$ through the hermitian matrix $O$ using Cholesky decomposition in combination with a shift-invert method \cite{Pietracaprina2018-supp}. This algorithm requires significantly less RAM compared to sparse LU-decomposition, at the cost of a less sparse matrix that needs to be diagonalized. Using this approach, the largest matrix size that we could diagonalize had Hilbert space dimensionalities equal to 32 for both the $a$- and $b$ modes. This corresponds to a matrix size of order $\approx10^6$, with a total RAM requirement of $\approx 1$ TB and $\approx 6$ hours of computation time per datapoint using the MATLAB software.

For the excited modes a trick like the one mentioned above will not work and one has to resort to using sparse LU-decompositions. Using this method we can compute the eigenvalues presented in the main text up to a cutoff of the Hilbert space dimension equal to 31. We note, however, that this is not enough to ensure high convergence for e.g. the period doubling modes for $N = 3$ and $4$. Nevertheless, we can qualitatively extrapolate the behavior of this system for higher values of $N$ as we did in the main text. Larger computational power, or more efficient schemes should be used to probe the period doubling behavior in more detail.

\subsection{Monte Carlo trajectories}
\begin{figure*}[t]
    \centering
    \includegraphics[width = \textwidth]{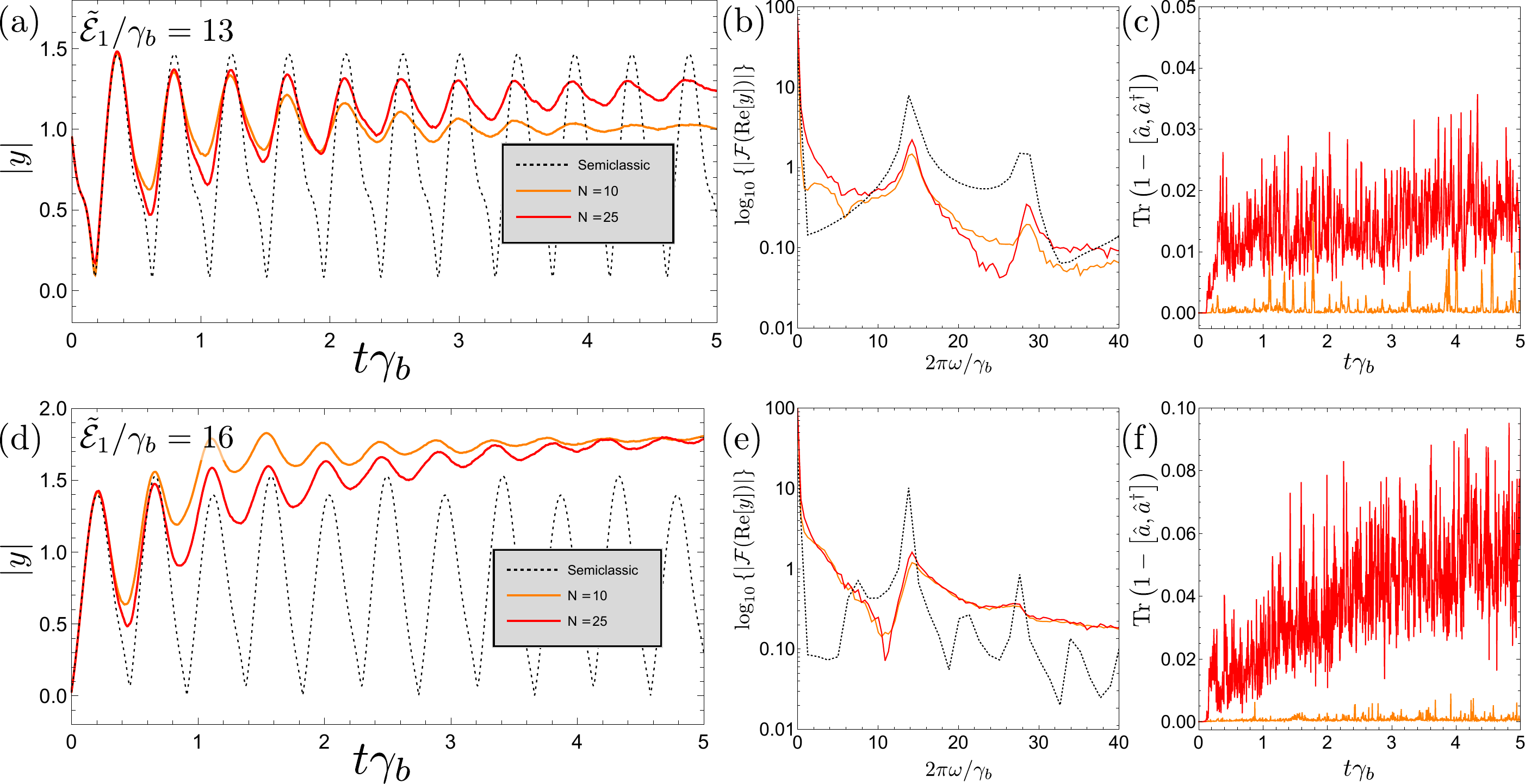}
    \caption{Monte Carlo trajectories of particle number expectation values $y_0 = \Tr [\hat{a}^\dag\hat{a}\rho(t)]$. Figs. (a), (b) and (c) show the averaged quantum trajectories at a driving amplitude of $\tilde{\mathcal{E}}_1/\gamma_b = 13$, the associated Fourier spectrum and error. The orange and red lines show the behavior for $N = 10$ and $N=25$ respectively with Fock-dimension cutoffs of $100$ and $250$. The semiclassical prediction is also included (dashed black lines). The initial state of the quantum trajectories is a coherent state, with an expectation value predicted by the semiclassical solutions. The quantum trajectories have been computed for a total time of $t_{\text{total}}\gamma_b = 15$, with $dt\gamma_b = 0.005$ and averaged over 3000 individual trajectories. Figs. (d), (e) and (f) show the same as the figures in the row above, but now at a driving amplitude of $\tilde{\mathcal{E}}_1/\gamma_b = 16$. Note that for this driving amplitude, the system decays faster to the steady state. This is expected from the larger dissipative gap at this parameter choice. The period doubled mode is visible in the semiclassical regime in the Fourier spectrum in figure (e). All parameters in the figure are the same as in the figures of main text.}
    \label{fig:supp_MonteCarloTraj}
\end{figure*}

In order to overcome the limitations of direct diagonalization we studied the quantum trajectories of the system using the QuTiP package for python \cite{Qutip2-supp}. This procedure is less sensitive to scaling of matrix sizes, and allows for larger system sizes up to $N = 25$ at a cutoff in Hilbert space dimensions (in the number basis) by 250. Results are shown in Fig.~\ref{fig:supp_MonteCarloTraj}. We clearly see that the trajectories decay to a steady state expectation value. This is coming from the fact that the dissipative gap is still finite. Increasing the parameter $N$ slows down the decay. From the Fourier spectrum we determine the periodicity of oscillations which closely match our semiclassical predictions. The period doubling mode remains elusive from these figures. This is due to the choice of the initial state. For the trajectories shown in Fig.~\ref{fig:supp_MonteCarloTraj}, we have chosen a coherent initial state on a point of the limit cycle predicted by the semiclassical analysis. This state has a negligible overlap with the period doubled mode, which can be deduced using the results of exact diagonalization methods. Therefore, the quantum trajectories are exceedingly unlikely to exhibit period doubling in their Fourier spectrum at the level of precision of the computations in this work.

\end{document}